\documentclass[aps, a4paper, reprint, twoside]{revtex4-1}

\usepackage[english]{babel}
\usepackage[utf8]{inputenc}
\usepackage[labelfont=up]{subcaption}
\usepackage[x11names]{xcolor}
\usepackage{float}
\usepackage[colorinlistoftodos, color=green!40, prependcaption]{todonotes}
\usepackage[pdftex, pdftitle={Article}, pdfauthor={Author}]{hyperref}
\hypersetup{colorlinks=true, citecolor=black, linkcolor=black, urlcolor=blue} 
\usepackage{mathtools}
\usepackage[top=1.in, bottom=1.in, left=0.5in, right=0.5in]{geometry}
\usepackage{amsthm, amssymb}
\usepackage{graphicx}
\usepackage{lipsum}

\DeclarePairedDelimiter\ton{(}{)}
\DeclarePairedDelimiter\qua{[}{]}
\DeclarePairedDelimiter\gra{\{}{\}}
\DeclarePairedDelimiter\mean{\langle}{\rangle}
\DeclarePairedDelimiter\av{\langle}{\rangle}
\newcommand{\me}{\mathrm{e}}
\newcommand{\erf}{\text{erf}}

\newcommand{\be}{\begin{equation}}
\newcommand{\ee}{\end{equation}}

\begin{document}

\title{Emergence of polarization in a voter model with personalized information}
\author{Giordano De Marzo}
\affiliation{Dipartimento di Fisica Universit\`a ``Sapienza”, P.le A. Moro, 2, I-00185 Rome, Italy.}

\author{Andrea Zaccaria}
\affiliation{Istituto dei Sistemi Complessi (ISC-CNR), UoS Sapienza,P.le A. Moro, 2, I-00185 Rome, Italy.}

\author{Claudio Castellano}
\email[Correspondence email address: ]{claudio.castellano@roma1.infn.it}
\affiliation{Istituto dei Sistemi Complessi (ISC-CNR), via dei Taurini 19, I-00185 Rome, Italy}

\date{\today} 
\begin{abstract}
The flourishing of fake news is favored by
recommendation algorithms of online social networks which, based on
previous users activity, provide content adapted to their
preferences and so create filter bubbles. 
We introduce an analytically tractable voter model with
personalized information, in which an external field tends to align the
agent opinion with the one she held more frequently in the past. Our
model shows a surprisingly rich dynamics despite its simplicity. An
analytical mean-field approach, confirmed by numerical simulations,
allows us to build a phase diagram and to predict if and how consensus is
reached. Remarkably, polarization can be avoided only for weak
interaction with the personalized information and if the number of
agents is below a threshold. We analytically compute this critical size,
which depends on the interaction probability in a strongly non linear way.
\end{abstract}

\maketitle
\section{Introduction}

The way our opinions are formed and change over time giving
rise to emerging collective phenomena is a topic that attracts
a rapidly increasing interest. Apart from its clear relevance
for fundamental issues, such as the stability of democracy and
the preservation of individual liberties,
opinion dynamics is a paradigmatic example of a social
phenomenon that can be quantitatively studied by a combination of 
theoretical and empirical 
approaches~\cite{lazer2009,castellano2009,acemoglu2011,Sen2014,Noorazar2020}.

Traditional models for opinion dynamics focus on the
interaction among a large number of peers, often in the presence
of external fields -- possibly varying over
time but equal for all agents~\cite{michard2005} --
describing the effect of conventional media, such as television and the press,
acting in the same way on all individuals.
The advent of the Internet, and the Online Social Network (OSN) revolution
in particular, have made the scenario more complex.
Although in principle the Internet allows users to access an unprecedented
diversity of news and viewpoints, this abundance is overwhelming and
pushes users to rely on automated recommendation systems to cope with
information overload. To provide successful recommendations web sites
and social networks constantly track our online activity and, based
on it, they propose personalized ``information'', which varies not only
over time, but also from person to person.
Examples of this phenomenon are posts appearing on Facebook news feed,
which are chosen and ordered by the social network on the basis
of previous interactions with other posts, or Google's Personalized PageRank.
The dependence on past user behavior coupled to the human inclination
to favor sources confirming one's own preferences~\cite{iyengar2009}
creates a feedback loop which drastically reduces the huge diversity of
available content.

The selective exposure of individuals to a biased representation
of the world (so that they remain confined within their
``filter-bubble'') is thought to play a crucial role in shaping
opinions at all scales~\cite{Bakshy2015}.
A most worrying aspect of this mechanism is the possible
reinforcement of personal biases
with the consequence of favoring radicalization phenomena~\cite{Maes2015}.
An astonishing example in this sense is the recent revelation that
64\% of people who join extremist groups on Facebook
are recommended to do so by the Facebook algorithm itself~\cite{Horwitz2020}.

It is therefore crucial to properly understand the effect of
personalized information or advertising on the dynamics of opinions.
Some modeling efforts in this direction have already been done.
Perra and Rocha~\cite{Perra2019} studied a binary opinion dynamics
where each user updates her opinion based on the opinion of
others filtered in various ways.
In the framework of continuous opinion dynamics with bounded
confidence~\cite{castellano2009}, in
Ref.~\cite{Sirbu2019} the effect of recommender algorithms in OSN
is mimicked by enhancing
the probability to interact with individuals having close opinions.
An increased tendency toward fragmentation (no consensus) and
polarization (clusters with distant opinions) is observed.
A model for polarization in continuous dynamics has been recently
presented and compared with empirical results by
Baumann et al.~\cite{Baumann2020}.

In this manuscript, we consider arguably the simplest possible
type of opinion dynamics, the voter model, and study in detail
its behavior in the presence of an external personalized information,
modeled, in its turn, in an extremely simple way.
The fundamental questions we want to answer are whether, in
this simple setting, selective exposure prevents the reaching
of consensus and how this comes about.
Is a minimal amount of personalized information sufficient to
lead to a polarized state? Is consensus possible for a very
strong influence of personalized information on opinions?
On which timescale is consensus, if any, reached?
What is the role of the system size?
By means of an analytical approach, corroborated by numerical
simulations, we fully understand the model behavior and thus
get complete answers to all these questions.

The rest of the paper is organized as follows.
In Section~\ref{secII} we introduce the voter model modified
by the presence of personalized information.
Its analytical investigation is described in Section~\ref{secIII},
divided in three subsections, dealing with different values of
the parameter $c$. The final Section summarizes and discusses the results
and presents some perspectives. Several Appendices contain details
of the analytical calculations.

\section{The voter model with personalized information}
\label{secII}

In voter dynamics~\cite{clifford1973, holley1975}
individuals are endowed with a binary opinion (spin)
$s_i = \pm 1$; at each time step a spin is randomly extracted and
its value is replaced with the value of one of the spins it is
connected to. In other words an individual becomes equal to a randomly
chosen neighbor.
The model has been studied extensively, both in regular lattices
and on complex networks~\cite{dornic2001, sood2005, castellano2009, Pugliese2009, Suchecki2005, FernandezGracia2014, Carro2016}.
It is well known that this dynamics always leads in finite systems
to full consensus -- i.e. all spins get aligned after a certain amount
of time -- and that this is due only to stochastic
fluctuations~\cite{Rednerbook}.

In order to understand the most basic effects of personalized
information on opinion dynamics, we couple the voter model
to a simple source of personalized information,
which feeds back on each individual a signal depending on the
past evolution of her own opinion.
Other recent works have investigated the effect of opposing
(but fixed) sources of external information on voter
dynamics~\cite{Bhat2019,Bhat2020}. Our model is also similar
to a voter dynamics with aging, recently introduced by
Peralta et al.~\cite{Peralta2020}. The main difference between
that model and the present one is that the effective memory
in our model is never erased, while it is, when a spin flips,
in the model of Ref.~\cite{Peralta2020}.

Let us consider $N$ agents distributed over the nodes $i$ of a network.
Each agent can assume two states $s_i=\pm 1$, that
correspond to two different opinions, and interacts with the agents it
is connected to. We define the adjacency matrix $A_{ij}$ so that
$A_{ij}=1$ if spin $i$ and $j$ are linked and
$A_{ij}=0$ otherwise. With this convention the number of agents a
given spin $s_i$ is connected to is simply $k_i = \sum_j A_{ij}$.
The evolution of each individual depends also on another variable,
a ``personalized external information'' $e_i$.
This last quantity is a random variable $e_i=\pm 1$ assuming the
positive value with a probability $P[e_i(t)=+1]$ that changes over time
depending on the history of the agent's opinion.

\begin{figure}
    \includegraphics[width=0.35\textwidth]{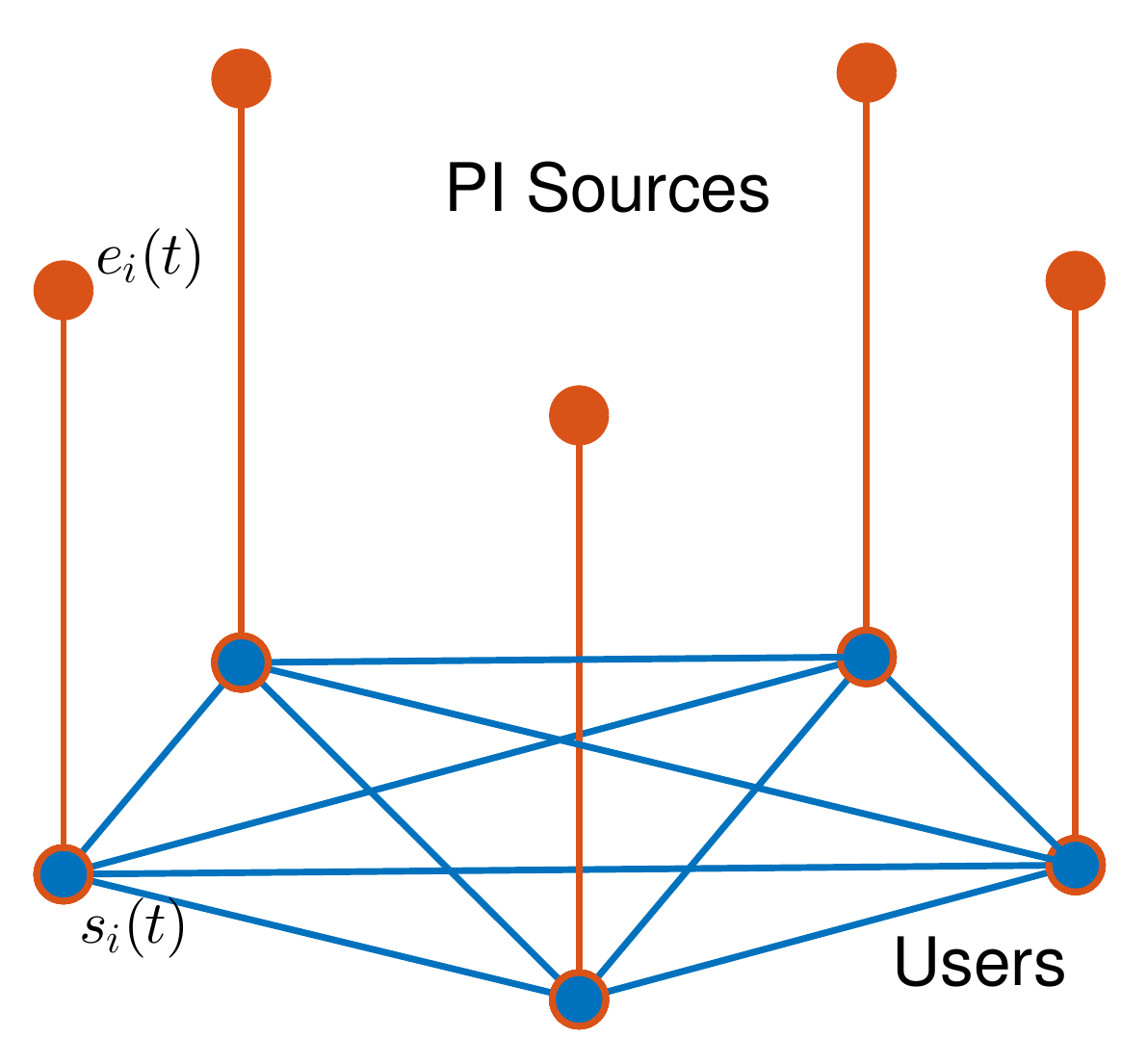}
    \caption{\textbf{Voter model with personalized information.}
      Graphical representation of the voter model with personalized
      information. Blue circles represent the agents $s_i$
      that interact following the usual voter dynamics, while red circles are
      the external agents, carrying personalized information. Note that the
      external agent $e_i$ influences only the corresponding voter
      spin, namely $s_i$.}
    \label{fig:vpiTxt}
\end{figure}

The dynamics takes place as follows.
Initially each spin is set to $s_i=\pm 1$ with equal probability.
At each time step, a given individual $i$ is selected at random and,
with probability $1-\lambda$, she follows the usual voter dynamics: Her
state $s_i$ is made equal to the state of a randomly selected neighbor
$s_j$.
With complementary probability $\lambda$,
the individual copies the state of the external source:
\begin{equation}
s_{i}(t+\delta t)=
\begin{cases}
e_i(t) \ \text{with probability}
\ \lambda \\ s_j(t) \ \text{with
  probability} \ \frac{1-\lambda}{k_i},
\end{cases}
\label{eq:voter}
\end{equation}
where $\delta t = 1/N$, $j$ is one of
the neighbors of $i$ (i.e., $A_{ij}=1$) and $N_i$ is the
total number of such neighbors.
Pictorially, we are adding another layer of ``external agents''
$e_{i}(t)$, each of them coupled only to the original agent
$s_{i}(t)$ and influencing her in the same way as the other neighbors,
except for a different probability of interaction. See Fig.~\ref{fig:vpiTxt} for a graphical representation of the model.

To mimic the reinforcing effect of personalized information we assume
that, whenever the spin $i$ is selected for update, then
$P[e_i(t)=+1]$ changes, increasing the probability that $e_i$
will be in the future equal to the current state of the agent, $s_i$.
More precisely at each time step, the update of the probability occurs after the update of the opinion variable. In other words, one first updates
the opinion variable (which may or may not change) and afterwards increases by a factor $c$ the ratio between $P[e_i=s_i(t+\delta t)]$ and $P[e_i=-s_i(t+\delta t)]$ where $s_i(t+\delta t)$ is the opinion variable after the update:
\[
\frac{P[e_i=s_i(t+\delta t)]}{P[e_i=-s_i(t+\delta t)]} \to 
c \frac{P[e_i=s_i(t+\delta t)]}{P[e_i=-s_i(t+\delta t)]}.
\] 
In this way the time-depending probability $P(e_i=+1)$
keeps track of the history of the spin $s_{i}$.
For example if agent $i$ stays in state $s_i=+1$ for a long time,
then $P(e_i=+1)$ tends to grow toward 1 and this makes more likely
that opinion $s_i=+1$ is maintained. The polarizing effect of this
personalized source of information is clear.
The parameter $c$ determines the speed at which the balance
between the two alternatives is disrupted.
Notice that the change for the probability $P(e_i=+1)$ occurs
at each update of agent $i$, even if the latter does not actually
change opinion (because the agent interacts with a neighbor already
sharing the same state).

It is useful to define the quantity
\[
n_i(t)=\sum_{t'=1}^t s_i(t'),
\]
which keeps memory of the evolution of agent $i$'s opinion.
Assuming that initially no knowledge about the agent's preferences
is available and therefore external information is fully
random $P[e_i(0)=+1]=1/2$ we can write
\be
P[e_{i}(t)=1] = P[n_i(t)]=\frac{c^{n_i(t)}}{1+c^{n_i(t)}}.
\label{eq:evolution_P(n_i)}
\ee
Hence, a positive (negative) value of $n_i$ implies that
personalized information is more probably equal to
$e_i=+1$ ($e_i=-1$).

\begin{figure*}
    \includegraphics[width=0.49\textwidth]{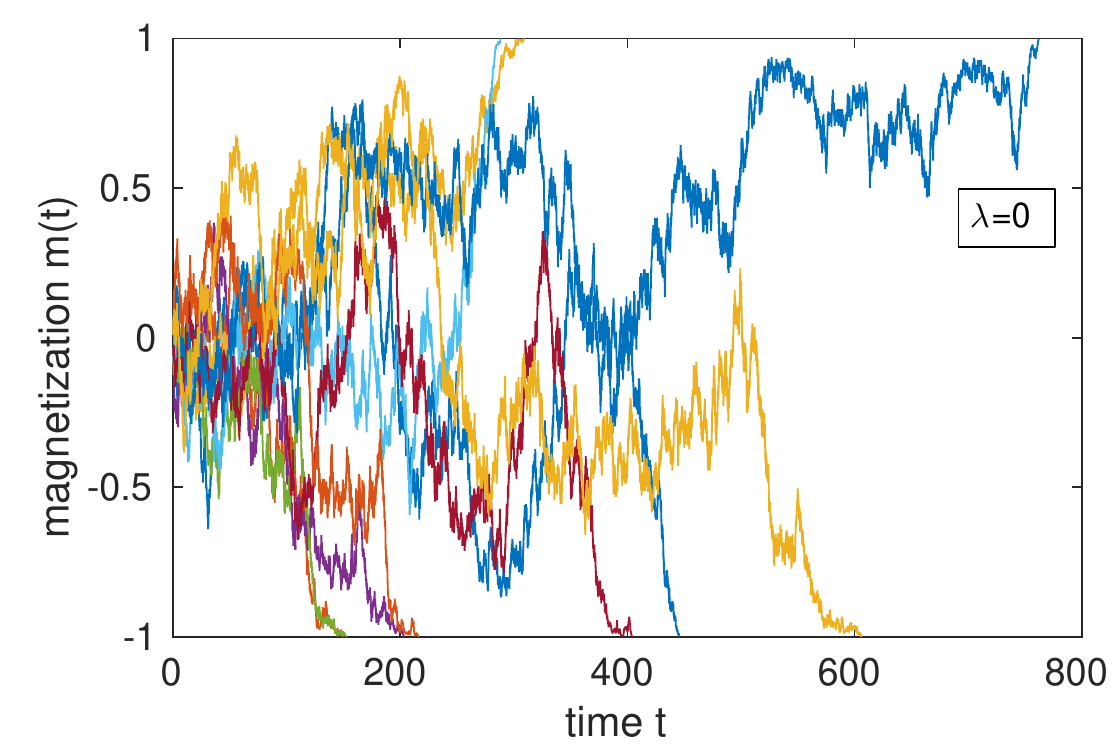}
    \includegraphics[width=0.49\textwidth]{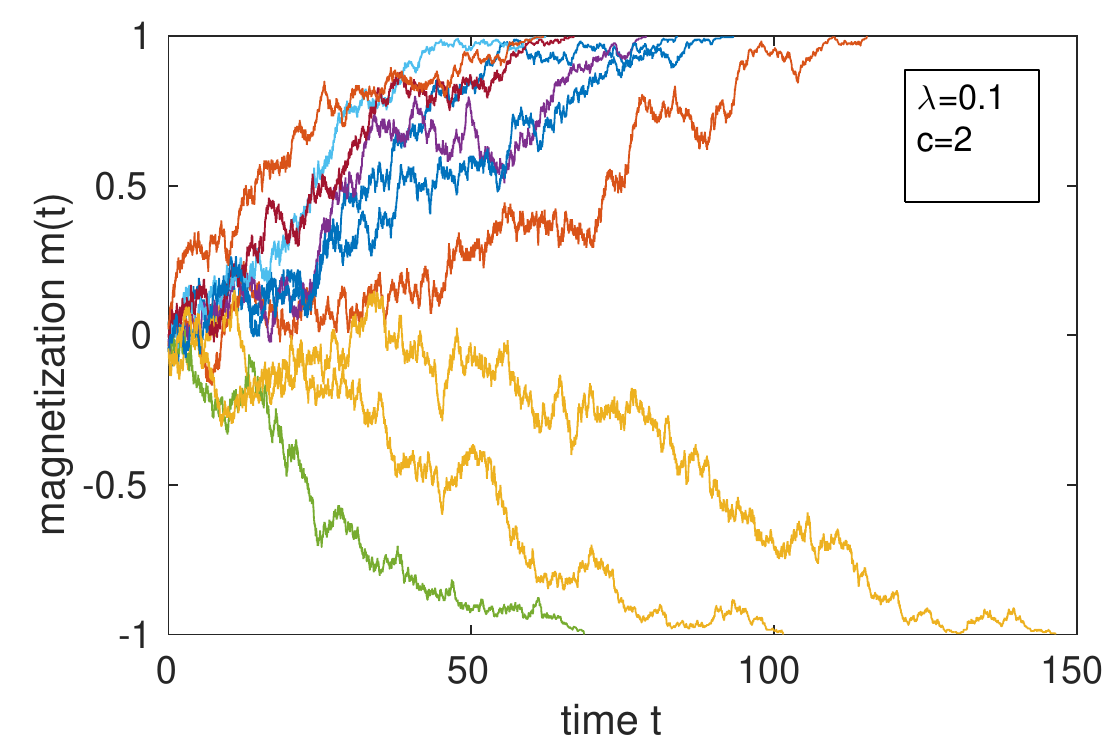}\\
    \includegraphics[width=0.49\textwidth]{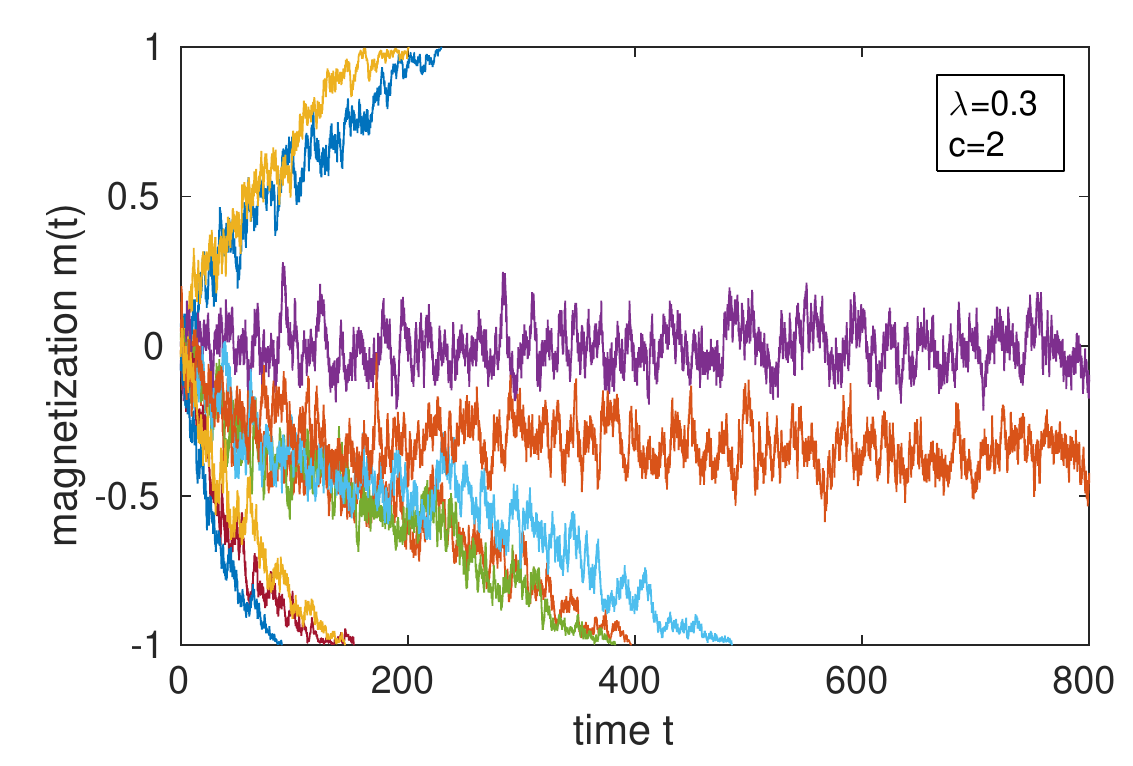}
    \includegraphics[width=0.49\textwidth]{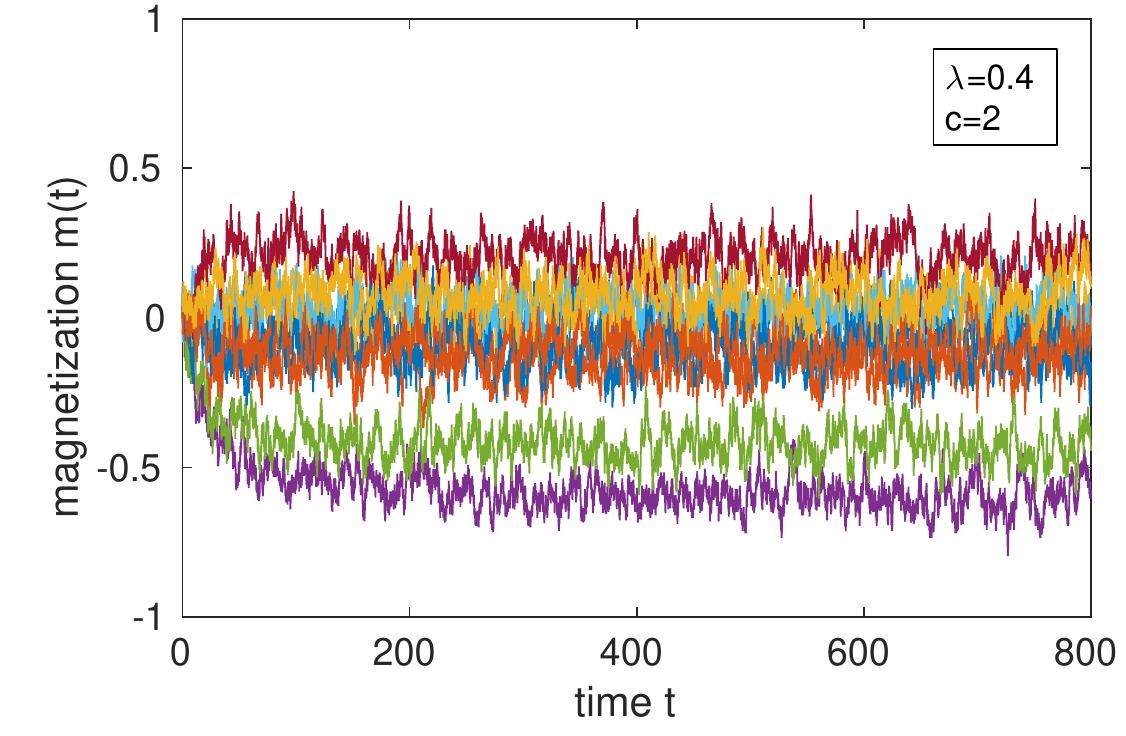}
  \caption{\textbf{From consensus to polarization.}
    Temporal evolution of the magnetization $m(t)$
    for different values of the probability $\lambda$,
    starting from $\lambda=0$ (pure voter
    dynamics). Agents form a complete graph of size $N=500$.
    The trajectories of 10 different runs are displayed.} 
  \label{fig:full}
\end{figure*}

To give a qualitative idea of the model behavior we report in
Fig.~\ref{fig:full} the temporal evolution of the magnetization
$m(t) = \sum_i s_i/N$ for different values of the probability
$\lambda$ of interaction with the personalized information.
For $\lambda=0$ the system is exactly the usual voter dynamics and
it reaches consensus $m = \pm 1$ because of random diffusive fluctuations,
in a time of order $N$~\cite{Rednerbook}.
As personalized information is turned on, consensus is still reached,
but surprisingly over {\em shorter} time intervals and
it is clear that drift plays now a relevant role.
Increasing $\lambda$ further we observe that some runs do not reach
consensus any more and magnetization fluctuates around some constant
value. Finally, for large $\lambda$ consensus is never reached and all
runs remain stuck in a disordered state.
In the next section, through a mean-field analytical approach,
we understand when and how consensus is reached,
depending on the values of the parameters
$c$ and $\lambda$.

\section{Analytical results}
\label{secIII}

The evolution of a system following Eqs.~\eqref{eq:voter} and
\eqref{eq:evolution_P(n_i)} depends on the topology of the
network defining the interactions among spins. In the following we
will focus on complete graphs, for which each pair of spins is
equally likely to interact, meaning that $A_{ij}=1-\delta_{ij}$ for
any $(i, j)$.
This corresponds, in the absence of external information,
to the mean field limit of the voter model.
We denote by $N_{\uparrow}$ the number of spins in state $+1$, while
$N_{\downarrow}=N-N_{\uparrow}$ is the number of spins in the opposite state.
In these terms the updating process is
\begin{equation}
  s_{i}(t+\delta t)=
  \begin{cases}
    e_i(t) \ \text{with probability} \ \lambda \\
    +1 \ \text{with probability} \ (1-\lambda)\frac{N_{\uparrow}}{N}\\
    -1 \ \text{with probability} \ (1-\lambda)\frac{N_{\downarrow}}{N},\\
  \end{cases}
  \label{eq:evolution_s_i_mean_field}
\end{equation}
where the probability of a positive external information
is given by Eq.~\eqref{eq:evolution_P(n_i)}.
Reminding that the magnetization can be written as
\[
m = \frac{N_{\uparrow}-N_{\downarrow}}{N},
\]
we obtain from Eq.~\eqref{eq:evolution_s_i_mean_field} that each time
a node is selected its value $s_i$ evolves according to
\begin{equation}
  s_i(t)\! \to \! s_i(t+\delta t)\!=\!\!
  \begin{cases}
    \!+1 ~~\textrm{w. prob:}~(1\!-\!\lambda)\!\left(\frac{1+m}{2}\right)+
    \lambda P(n_i) \\
    \!-1 ~~\textrm{w. prob:}~(1\!-\!\lambda)\!\left(\frac{1-m}{2}\right)+
    \lambda [1\!-\!P(n_i)].
  \end{cases}
  \label{eq:evolution_s_i}
\end{equation}
and, immediately after, $n_i$ is updated as follows
\begin{equation}
  n_i(t) \to n_i(t+\delta t)=
  \begin{cases}
    n_i(t)+1 ~~~\textrm{if}~~~s_i(t+\delta t)=1 \\
    n_i(t)-1 ~~~\textrm{if}~~~s_i(t+\delta t)=-1.
  \end{cases}
  \label{eq:evolution_n_i}
\end{equation}
Thus the state of each node is defined by the pair
$(s_i, n_i)$ and therefore the evolution of the system depends on the set
$\gra*{(s_i, n_i)}_{i=1}^N$. In Appendix~\ref{coefficients}
we calculate the drift and diffusion coefficients~\cite{Rednerbook}
for the magnetization and for the average value of the $n_i$
\be		
n=\frac{1}{N} \sum_i n_i.
\label{n}
\ee
We obtain the magnetization drift
\be
  v^m =
2 \frac{\lambda}{N}\sum_i \left\{ \left(\frac{1-s_i}{2} \right) P(n_i) - \left(\frac{1+s_i}{2} \right) [1-P(n_i)] \right\} 
  \label{eq:drift_m_general}
  \ee
and the magnetization diffusion coefficient
\begin{eqnarray}
  \label{eq:diffusion_m_general}
  D^m &=& \frac{1-\lambda}{N}(1-m^2) \\ \nonumber
  &+& \frac{2\lambda}{N^2}
  \sum_i \left\{ \left(\frac{1-s_i}{2} \right) P(n_i) + \left(\frac{1+s_i}{2} \right) [1-P(n_i)] \right \}.
\end{eqnarray}
For the quantity $n$ the drift coefficient reads
\be
v^n = (1-\lambda) m + \frac{\lambda}{N} \sum_i \left[2 P(n_i)-1 \right].
\label{eq:drift_n_general}
\ee
while the diffusion coefficient is
\begin{equation}
  D^n = \frac{1}{N}.
  \label{eq:diffusion_n_general}
\end{equation}

These expressions contain $P(n_i)$ and therefore are
different depending on the value of $c$.

\subsection{The case $c=1$}

Let us first discuss the case $c=1$, that is equivalent to the noisy voter model or Kirman
  model \cite{kirman1993ants}.
In this case Eq.~\eqref{eq:evolution_P(n_i)} reduces to
\[
P(n_i)=\frac{1}{2} \ \forall \ i,t.
\]
and therefore the variables $n_i$ do not play any role.
Setting $c=1$, Eqs.~\eqref{eq:drift_m_general} and
\eqref{eq:diffusion_m_general} reduce to
\begin{equation}
  \begin{cases}
    v^m_{c=1}=-\lambda m \\
    D^m_{c=1}=\frac{1}{N}\qua*{\ton*{1-\lambda}\ton*{1-m^2}+2\lambda}
  \end{cases}
  \label{eq:nu_D_c=1}
\end{equation}
Differently from the standard voter model there is a nonzero drift
term, driving the system toward the disordered symmetric configuration
$m=0$. However, depending on the value of $\lambda$, the
  system may still spend most of its time in the consensus state
  $m=\pm 1$, that is no more absorbing.
  See Refs.~\cite{Alfarano2005, artime2018aging, artime2018first}
  for a detailed analysis of the noisy voter model.

\subsection{The behavior for $c \gtrsim 1$}

We now study the behavior for $c>1$, considering separately two cases.
We set $c=1+\delta$ and first take $\delta\ll 1$.
Under this hypothesis and focusing on short times
we can expand $P(n_i)$ to first order in $n_i \delta$, obtaining
\be
P(n_i)=\frac{c^{n_i}}{1+c^{n_i}}=\frac{(1+\delta)^{n_i}}{1+(1+\delta)^{n_i}}\approx\frac{1}{2}+\frac{n_i\delta}{4}.
\ee
Inserting this expression into Eq.~\eqref{eq:drift_m_general}
we get
\begin{equation}
  v^m \approx \lambda \left(\frac{n \delta}{2} - m \right),
  \label{eq:drift_m_delta}
\end{equation}
and analogously for the diffusion coefficient
\begin{equation}
  D^m\approx \frac{(1-\lambda)}{N} (1-m^2) + \frac{2 \lambda}{N^2}
  \ton*{1-\frac{\delta}{2N}\sum_i s_i n_i }.
  \label{eq:diffusion_m_delta}
\end{equation}

Inserting the expansion of $P(n_i)$ into Eq.~(\ref{eq:drift_n_general})
we obtain
\begin{equation}
  v^n \approx
  (1-\lambda) m + \frac{\lambda \delta}{2} n.
  \label{eq:drift_n_delta}
\end{equation}
			
\begin{figure}
  \includegraphics[width=0.45\textwidth]{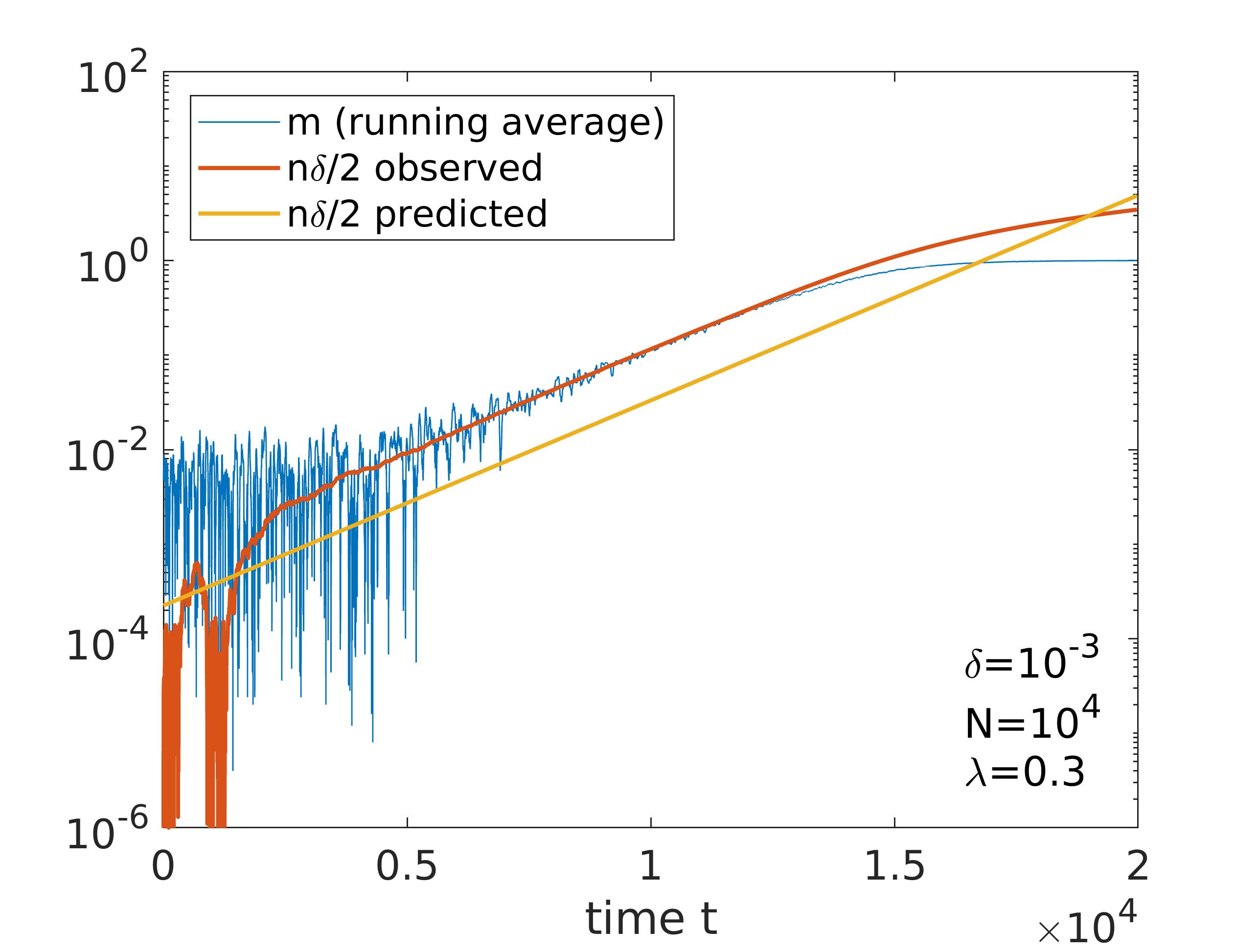}
  \caption{\textbf{Growth of $n$ and $m$.} Growth of $m$ compared with
    the observed $n\delta/2$ and with the predicted trend, as defined
    by Eq.~\eqref{eq:m_n_delta}. $n$ grows diffusively up to
    $\tau=2/\delta$, time at which the exponential contribution
    becomes dominant. }
  \label{fig:m_n}
\end{figure}

In summary, combining Eqs.~\eqref{eq:drift_m_delta} and \eqref{eq:drift_n_delta}, the evolution of the system is given, as long as the condition
$|n_i|\delta \ll 1$ is satisfied for any $i$, by
\begin{equation}
  \begin{cases}
    \dot{m} = - \lambda m + \frac{\lambda \delta}{2} n \\
    \dot{n} = (1-\lambda) m + \frac{\lambda \delta}{2} n,
  \end{cases}
  \label{eq:system_m_n_delta}
\end{equation}
where fluctuations due to diffusion have been neglected.
By integrating we find, under the assumption $\delta \ll \lambda$,
\begin{equation}
\begin{cases}
  m = C_1 \frac{\delta}{2} \me^{t\delta /2} - C_2 \frac{\lambda}{1-\lambda}\me^{-\lambda t}, \\
  n = C_1 \me^{t\delta/2} + C_2 \me^{-\lambda t},
\end{cases}
  \label{mn}
\end{equation}
where $C_1$ and $C_2$ are determined by the initial conditions.  We
note that the deterministic evolution described by
Eq.~(\ref{eq:system_m_n_delta}) is preceded by a regime dominated by
stochastic effects, where we can effectively assume $c=1$.  During
such an interval $m$ fluctuates around zero due to the presence of the
term $-\lambda m$ in its drift, with fluctuations of the order of
$\pm m_0 = \pm \sqrt{D^m} \approx \pm \sqrt{(1-\lambda)/N}$.
Conversely $n$ grows diffusively, up to the
time $\tau=2/\delta$ after which the exponential growth becomes
dominant. Moreover, after a short time of order $1/\lambda$ the terms
proportional to $C_2$ in Eqs.~(\ref{mn}) become negligible. As a
consequence we can use as initial condition for $n$ its value at time
$\tau$, i.e., at the end of the diffusive regime, yielding
$C_1\approx\sqrt{\tau D_n}\approx\sqrt{\frac{2}{\delta N}}$
\begin{equation}
  \begin{cases}
    n \approx \pm \sqrt{\frac{2}{\delta N}}\me^{t \delta/2}\\
    m=\frac{\delta}{2}n.
  \end{cases}
  \label{eq:m_n_delta}	
  \text{for}\ t>\tau  	
\end{equation}
Note that the exponential growth of $m$ actually begins only when
$\frac{\delta}{2}|n| \sim m_0 \approx\sqrt{(1-\lambda)/N}$.
Fig.~\ref{fig:m_n} shows that Eq.~\eqref{eq:m_n_delta} describes well
this stage of the temporal evolution of $n$ and $m$.

The linear approximation remains valid until the time $T_c$ when,
for some $i$, $|n_i|$ becomes so large that the condition
$|n_i| \delta \ll 1$ breaks down.
This may happen in two different ways, depending on the shape of the probability distribution $Q(n_i)$ of the variables $n_i$.
  When the linear approximation breaks down, if the standard deviation of
  $Q(n_i)$ is small, the personalized information is approximately the same
  for all individuals. As a consequence this
  source of information can be regarded as a constant field and this
  results in the presence of a drift for the magnetization, which
  fastly reaches $m=\pm1$. Conversely, if the standard deviation is
  much larger than the average $\langle n_i\rangle$, some of the spins
  are characterized by a negative $n_i$, while $n_i$ is positive for the 
  others. This implies that a fraction of the spins is influenced by a
  positive external field, while a negative field acts on the
  remaining part, thus leading to a polarized state. More in detail,
if $Q(n_i)$ is narrowly peaked around its mean value $n$ over time,
linearization breaks down for a time $T$ such that $|n(T)| \delta \sim
1$.  From Eq.~\eqref{eq:m_n_delta} \be
T=\frac{1}{\delta}\log\ton*{\frac{N}{2\delta}}.  \ee At this time all
$n_i$ have the same sign, hence the drift (see Appendix~\ref{coefficients})
\be v_i = m(T) (1-\lambda) +
\lambda[2P(n_i)-1],
\ee
has the same sign for all individuals and thus
consensus is rapidly reached.

\begin{figure*}
  \includegraphics[width=0.9\textwidth]{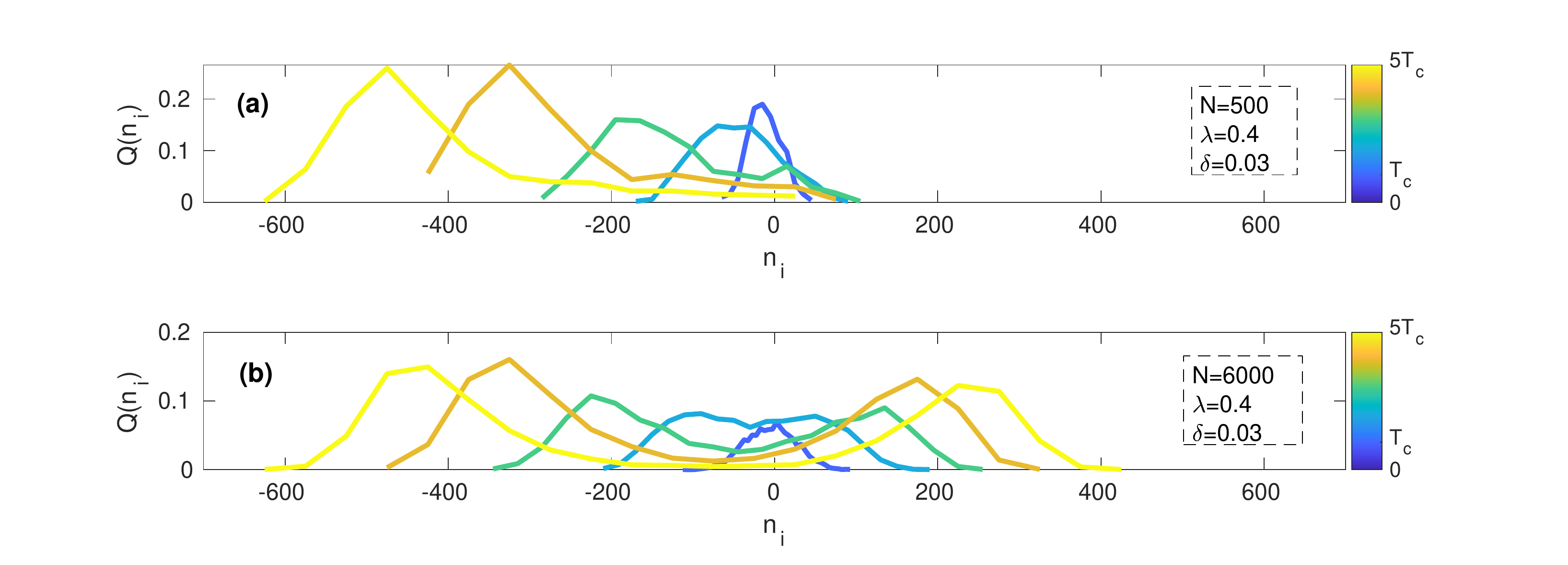}
  \caption{\textbf{Evolution of $Q(n_i)$.} Evolution of $Q(n_i)$
    for $N<N_c$ (panel (a)) and $N>N_c$ (panel (b)).}     
  \label{fig:N_c_bimodal}
\end{figure*}

Alternatively, if the distribution is very broad
so that its standard deviation $\sigma$ is much larger than the
absolute mean value $|n|$, linearization starts to fail at a different
time $T^*$.
Assuming, for the sake of simplicity, $n>0$, this occurs when
\be
\delta [n + \sigma(T^*)] \approx \delta \sigma(T^*)=1
\ee
The calculation of the variance $\sigma^2(t)$ of the distribution $Q(n_i)$,
reported in Appendix~\ref{Smallc}, gives
\begin{equation}
  \sigma^2(t) = \frac{1}{\lambda \delta} \left(\me^{\lambda \delta t} - 1\right).
  \label{eq:variance_delta}
\end{equation}
Imposing $\delta \sigma(T^*) = 1$
we then obtain
\be
T^*=\frac{1}{\lambda\delta}\log\ton*{1+\frac{\lambda}{\delta}}.
\label{Tstar}
\ee
The time $T_c$ when the linear approximation breaks down consequently is
\[
T_c=\min(T, T^*).
\]
Since $T$ grows with $N$, while $T^*$ does not depend on it, for small size
$T<T^*$ and the opposite relationship $T>T^*$ is instead true for large $N$.
Setting $T=T^*$ we can compute the crossover size
\be
N^*=2\delta\ton*{1+\frac{\lambda}{\delta}}^{1/\lambda}.
\label{eq:N^*}
\ee
For $N<N^*$ linearization breaks down due to the growth of $n$ and
as a consequence consensus is always reached, all the drifts having
the same sign.  Differently, if $N>N^*$ the end of the linear regime
is caused by the growth of the variance.  In this second case, at
$T^*$ most of the individuals have positive $n_i$ and hence a positive
drift, but some have negative $n_i$ (see Fig.~\ref{fig:N_c_bimodal}).
Determining in this case whether consensus is reached
or not is more involved, as discussed in the following.

\begin{figure}
  \includegraphics[width=0.49\textwidth]{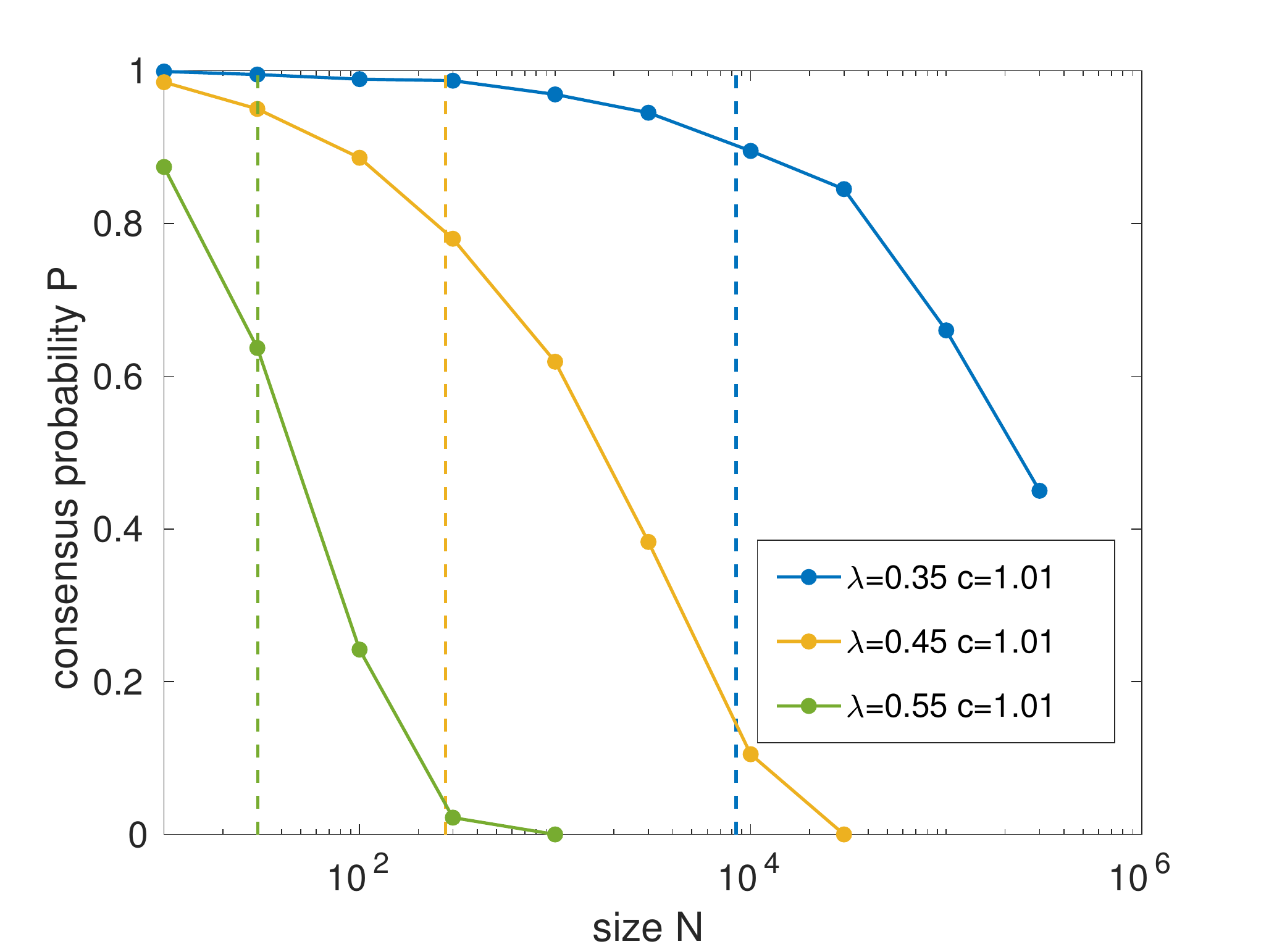}
  \caption{\textbf{Crossover between consensus and polarization for small $c$.}
      Fraction of
      runs reaching consensus as a function of $N$,
      in 1000 realizations of the dynamics for $\delta=0.01$
      and various $\lambda$. The crossover values $N_c$ predicted by
      Eq.~(\ref{N_csmallc}) are marked by vertical lines.}
  \label{pdchecksmallc}
\end{figure}

Assuming $N\gg N^*$, so that the standard deviation is much larger than the
mean value, the smallest of the negative values is
\be
n_i \approx n-\sigma(T^*) \approx - \sigma(T^*) \approx - \frac{1}{\delta},
\ee
and as a consequence the smallest drift is
\be
v_i = m(T^*) (1-\lambda) +
\lambda [2 P(n_i)-1] \approx m(T^*) (1-\lambda) -
\lambda.
\label{eq:drift_v_i}
\ee
If this value is positive, the corresponding individual,
which is the one whose external information is more negatively polarized,
will be pushed towards positive values of $n_i$. It then
follows that also in this case the system reaches consensus.
More quantitatively, the condition for having consensus, $v_i>0$,
implies (considering also the symmetric case when $n<0$)
\be
|m(T^*)| > \frac{\lambda}{1-\lambda}=m_c.
\label{mstar}
\ee
Inserting Eq.~(\ref{Tstar}) into Eq.~(\ref{eq:m_n_delta}) we obtain
\be
|m(T^*)| = 
\sqrt{\frac{\delta}{2N}} \ton*{1+\frac{\lambda}{\delta}}^{\frac{1}{2 \lambda}},
\ee
that combined with the condition~(\ref{mstar}) implies
that consensus is undoubtedly reached for $\lambda<1/2$ if $N<\bar{N}$, with
\be
\bar{N} = \frac{(1-\lambda)^2}{\lambda^2}\frac{\delta}{2}
\ton*{1+\frac{\lambda}{\delta}}^\frac{1}{\lambda}.
\label{bar_N_smallc}
\ee
Note, however, that this is only a lower bound for the true $N_c$ for
$\lambda<1/2$.
Actually consensus may occur also for $N>\bar{N}$, provided that $\lambda<1/2$.
Indeed, even if the smallest
drift (corresponding to the most negative $n_i$) is negative when
linearization breaks down, it can become positive later on, thus making the
system reach consensus. This may occur if only a few spins (among those
with $n_i<0$) have a negative drift. The others, moving toward
positive $n_i$, produce an increase of the magnetization, which
eventually overcomes the critical value $m_c=\frac{\lambda}{1-\lambda}$.
The critical size $N_c$, determining if the system can reach consensus
or not, is then larger than $\bar{N}$ for $\lambda<1/2$.
It is actually possible to improve on this result.
In Appendix~\ref{refined} a more refined argument is presented, allowing
us to determine numerically a tighter lower bound $\hat{N}$ for $N_c$.
This bound, as shown in Fig.~\ref{pdchecksmallc}, is in good agreement with simulations.

The situation is different if $\lambda\geq 1/2$, because in this case
the critical magnetization is larger than 1 and therefore,
even if some spins move from negative
$n_i$ to positive $n_i$, the smallest drift remains negative,
for any magnetization.
This implies that for $\lambda\geq 1/2$,
consensus can be reached only if linearization breaks down due to the
growth of $n$.
Therefore if $\lambda\geq1/2$,
the critical size coincides with $N^*$.
In conclusion
\be
\begin{cases}
  N_c \geq \hat{N}~~~~~~~~~~~~~~~~~~~~~~~~ \text{for}\ \lambda<\frac{1}{2}\\
  N_c=N^*=2\delta\ton*{1+\frac{\lambda}{\delta}}^{1/\lambda}\ \text{for}\ \lambda\geq\frac{1}{2}.
\end{cases}
\label{N_csmallc}
\ee
Of course, given the dependence of the argument on random fluctuations,
these values are to be intended as indicating a crossover and not a sharp
transition. Simulation results presented in Fig.~\ref{pdchecksmallc}
confirm that the probability of reaching consensus exhibits for various
$\lambda$, a crossover in reasonable agreement with Eq.~\eqref{N_csmallc}.

\subsection{The behavior for $c \gg 1$}	
\begin{figure*}
  \begin{subfigure}[b]{0.475\textwidth}
    \includegraphics[width=0.9\textwidth]{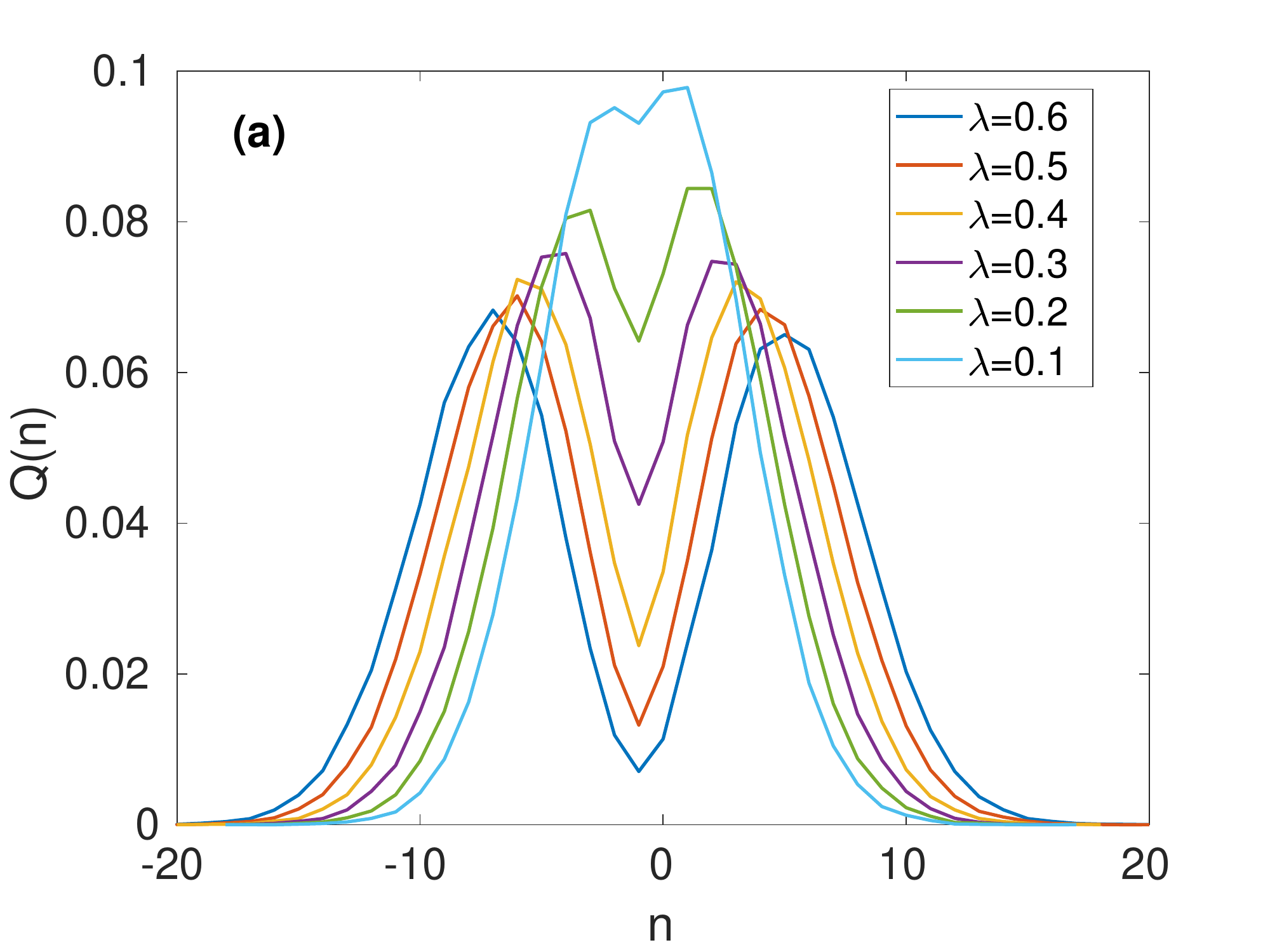}     
    \label{fig:distribution_lambda}
  \end{subfigure}
  \hfill
  \begin{subfigure}[b]{0.475\textwidth}  
    \includegraphics[width=0.9\textwidth]{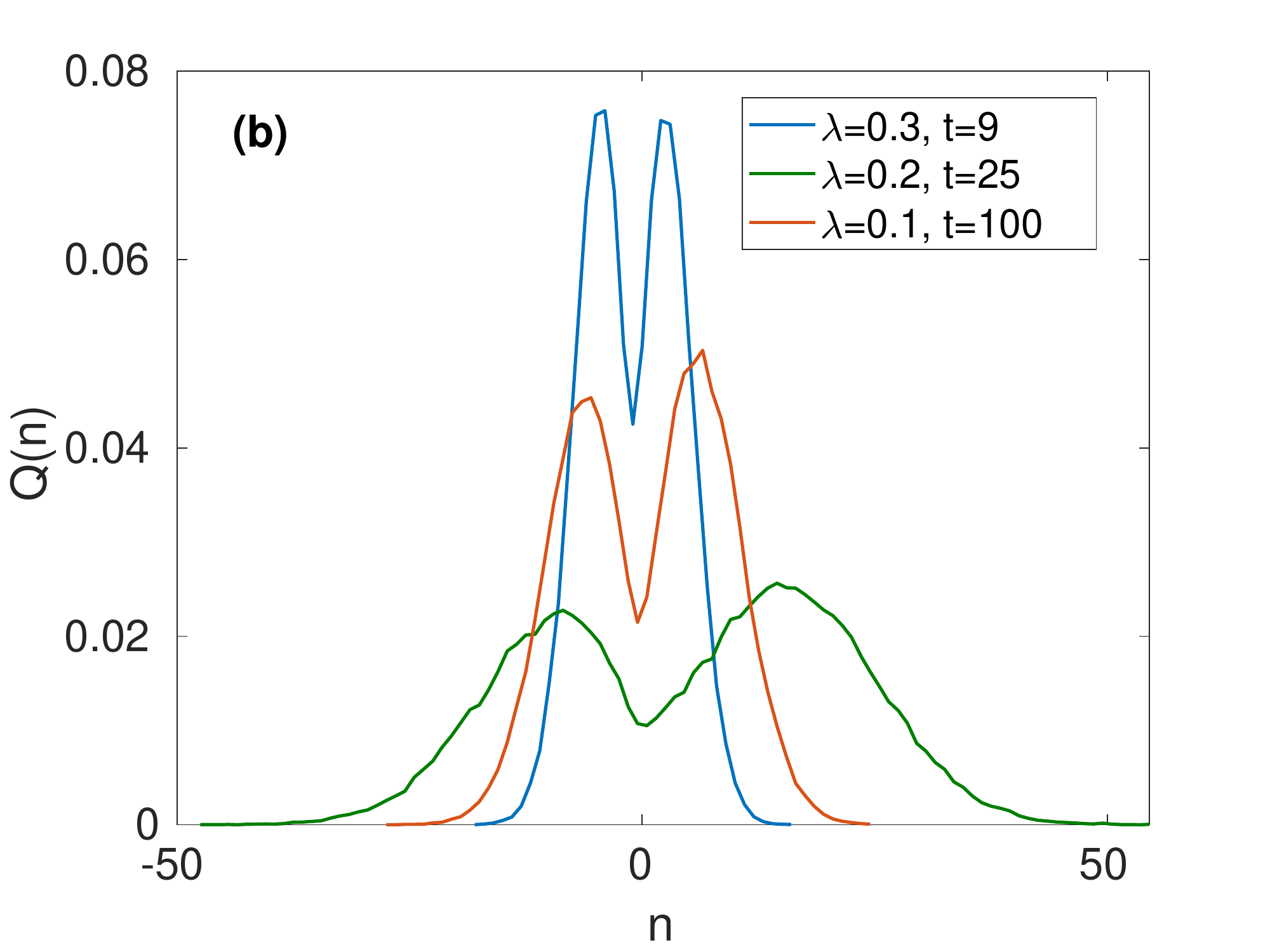} 
    \label{fig:distribution_time}
  \end{subfigure}\\
  \caption{\textbf{Splitting of $Q(n_i)$.} \textbf{a)} Distribution
    $Q(n_i)$ at $t=10$ for various values of $\lambda$. Note that, for
    $\lambda=0.1$, $t \ll t_c = 100$ while for $\lambda=0.6$ $t \gg
    t_c = 2.77$.  \textbf{b)} Distribution $Q(n_i)$ at $t \approx t_c
    = 1/\lambda^2$ for various values of $\lambda$.}
    \label{fig:distributions}
\end{figure*}
As shown in the previous subsection, if $c$ is close to $1$ the system
can be described in terms of the magnetization and of the
first two moments of the distribution $Q(n_i)$, $n$ and $\sigma^2$.
This is possible because (see Fig.~\ref{fig:N_c_bimodal})
this distribution is unimodal during the linear regime,
that lasts for long times, up to $T_c=\min(T,T^*)$.
Conversely, when $c$ is large, the pair $(n,\sigma^2)$ alone
is not sufficient to describe the state of the system, even in the
first stage of the dynamics.
This can be seen by inspecting the local probabilities $P(n_i)$.
If $c\gg 1$ we have
\[
P(n_i)=
\begin{cases}
  1-c^{-n_i} \approx 1 \ ~~ \text{if} \ n_i>0 \\
  \frac{1}{2} \ ~~~~~~~~~~~~~~~~ \text{if} \ n_i=0 \\
  c^{n} \approx 0 \ ~~~~~~~~~~ \text{if} \ n_i<0.
\end{cases}
\]
This means that, as soon as $n_i$ becomes different from zero at
the first update of node $i$, the
external information almost certainly has the same
sign of $n_i$ and thus tends to increase its absolute value.
The behavior of nodes with different signs of $n_i$ is
opposite and this rapidly leads to the splitting of $Q(n_i)$ in
two separate peaks.
We then write the overall distribution as the sum of two unimodal
distributions 
\[
Q(n_i)=pQ^+(n_i)+(1-p)Q^-(n_i).
\]
Here $Q^+(n_i)$ is the distribution of positive $n_i$,
while $Q^-(n_i)$ is the distribution of the negative ones and the
weight is $p=N^+/N$.
By using the same formalism described above, 
in Appendix~\ref{Largec} we derive expressions for the moments
of the two distributions
\be
v^\pm = \frac{d\av{n_i}_\pm}{dt} = m(1-\lambda) \pm \lambda,
\label{vpm}
\ee
\be
\sigma^2_{\pm} = t.
\ee
Since $m$ is initially very small, Eq.~(\ref{vpm}) confirms
that positive (negative) $n_i$ tend to increase (decrease)
and the overall distribution $Q(n_i)$ broadens and eventually splits.
In the same Appendix we derive the expression for
the variance $\sigma^2(t) = \av{n_i^2}-\av{n_i}^2$
of the whole distribution $Q(n_i)$, finding
\begin{equation}
  \sigma^2(t)=t+\lambda^2t^2.
  \label{eq:sigma(t)}
\end{equation}
From Eq.~\eqref{eq:sigma(t)} we can determine the time at which
$Q(n_i)$ splits in two separate components.
For small times $\sigma(t)$ is dominated by
the diffusive widening of the central peak,
while for larger times it is determined mainly by the ballistic
distancing between the two peaks. Denoting by $t_c$ the time at which
the crossover occurs, it follows from Eq.~\eqref{eq:sigma(t)}
$t_c\approx\lambda^2 t_c^2$
yielding
\[
t_c\approx\frac{1}{\lambda^2}.
\]
Figures~\ref{fig:distributions}a and~\ref{fig:distributions}b
confirm numerically this prediction.
Note that, at odds with the case $c \gtrsim 1$, the splitting
always occurs, and over a much shorter temporal scale since
$t_c = 1/\lambda^2 \ll \log(1+\lambda/\delta)/(\lambda \delta) = T^*$.

To understand whether the system reaches consensus or not, the
argument is similar to the one presented for $c \gtrsim 1$, but in
this case it provides the actual critical size $N_c$ rather than a
lower bound.  If, when the distribution splits, the drift of the right
peak is positive and that of the left peak is negative, consensus is
not reached.  From Eq.~(\ref{vpm}) this implies that consensus is
reached only if
\be
|m(t_c)| > m_c=\frac{\lambda}{1-\lambda}.
\label{condition}
\ee

This condition means, as before, that consensus for $\lambda>1/2$
can occur only before the splitting, so during the transient regime.
It turns out numerically that, for $t<t_c$, the magnetization $m$ grows as
$m(t) \approx \xi t/\sqrt{N}$, where $\xi$ is a random prefactor ranging
between approximately -1 and +1.
Hence for $\lambda>1/2$ the condition for consensus reads 
\[
|m(t)|=1 \ \text{for}\ t<t_c\ \to \ \frac{t_c}{\sqrt{N_c}}=1\ \to\ N_c=\frac{1}{\lambda^4}.
\]

For $\lambda<1/2$ instead, inserting the expression for $m(t)$ into
Eq.~(\ref{condition}) yields that consensus cannot be reached
if the number $N$ of individuals is larger than
\be
N_c =
\frac{(1-\lambda)^2}{\lambda^6}.
\label{N_c}
\ee

For $N>N_c$ the system remains asymptotically disordered in a polarized
state.
In the opposite case instead consensus is rapidly reached after
$t_c$, unless by chance the initial absolute value of $\xi$
is particularly small.

\begin{figure}
  \includegraphics[width=0.49\textwidth]{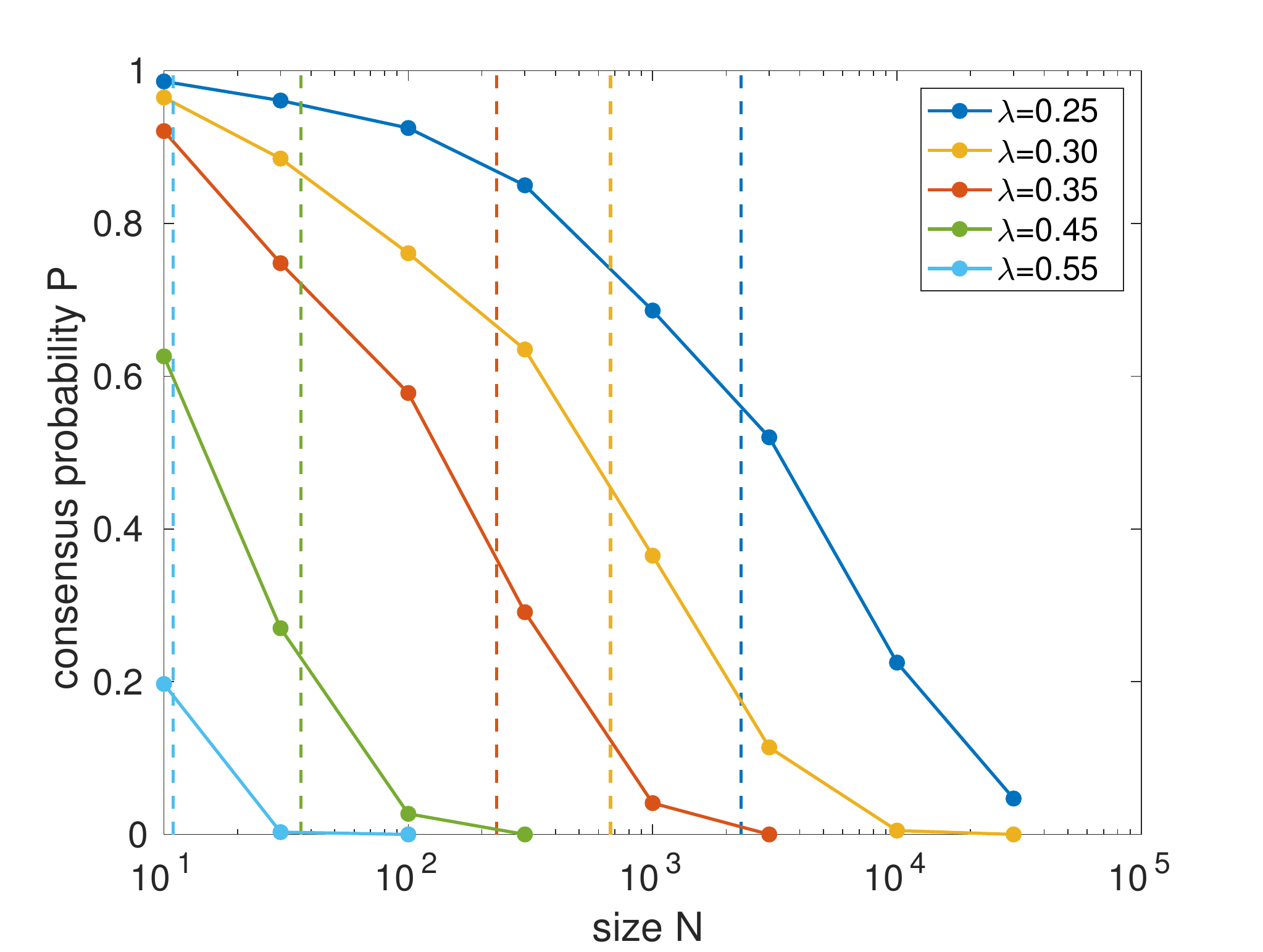}
  \caption{\textbf{Crossover between consensus and polarization for large $c$.}
      Fraction of
      runs reaching consensus as a function of $N$,
      in 1000 realizations of the dynamics for $c=100$
      and several $\lambda$ values. The crossover values $N_c$ predicted by
      Eq.~(\ref{N_c2}) are marked by vertical lines.}
  \label{pdcheck}
\end{figure}

In conclusion, recalling Eq.~\eqref{N_c}, the critical size satisfies
\be
\begin{cases}
  N_c = \frac{(1-\lambda)^2}{\lambda^6}~~~ \text{for}\ \lambda<\frac{1}{2}\\
  N_c = \frac{1}{\lambda^4}~~~~~~~~ \text{for}\ \lambda\geq\frac{1}{2}.
\end{cases}
\label{N_c2}
\ee
Note that $N_c$ is continuous in $\lambda=1/2$.
Simulations presented in Fig.~\ref{pdcheck} show that the probability
of reaching consensus exhibits a crossover at values well predicted
by Eq.~(\ref{N_c2}).

\begin{figure*}
  \begin{subfigure}[b]{0.475\textwidth}  
    \includegraphics[width=0.91\textwidth]{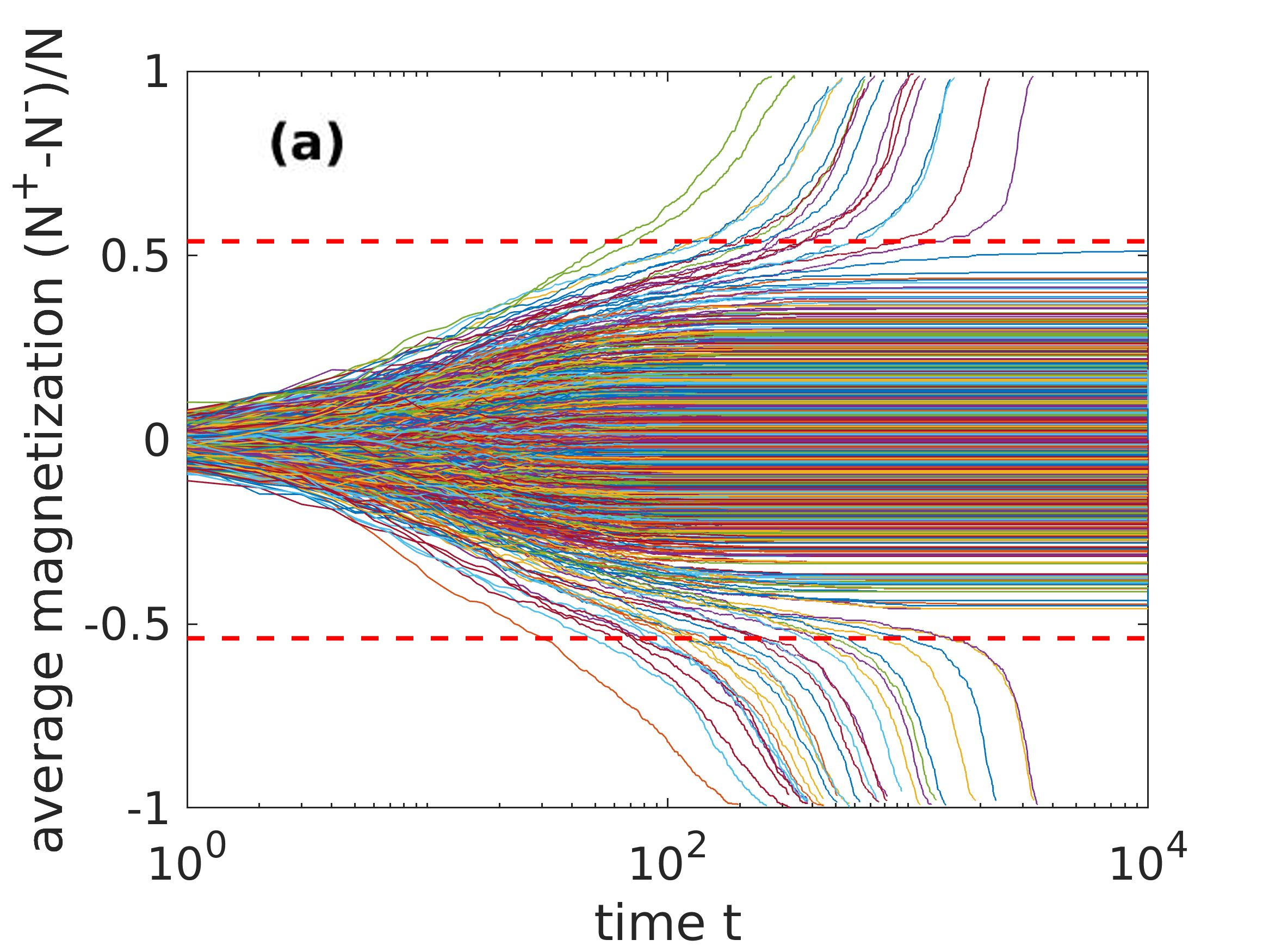}   
    \label{fig:many_trajectories}
  \end{subfigure}
  \hfill
  \begin{subfigure}[b]{0.475\textwidth}
    \includegraphics[width=0.95\textwidth]{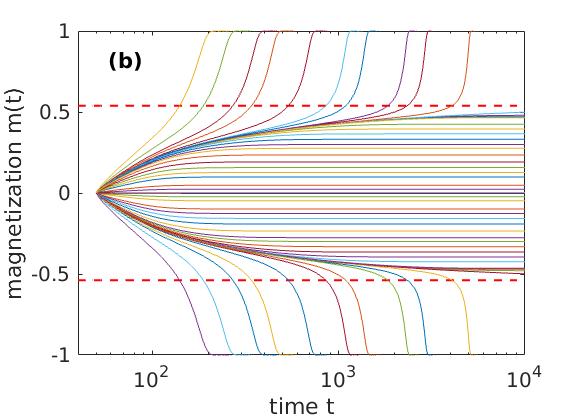}
    \label{fig:MagnMcEqDiff}
  \end{subfigure}
  \caption{\textbf{Time evolution of the magnetization for $c=100$.}
    \textbf{(a)} $10^4$ trajectories obtained simulating the  model
    with $N=10^3$, $c=100$ and $\lambda=0.35$. Red dashed lines
    represent $\pm m_c=\pm\lambda/(1-\lambda)$, disordered states are
    found in the region $[-m_c;+m_c]$. \textbf{(b)} Solutions of
    Eq.~\eqref{integrodiff} for $\lambda=0.35$ and various $m_0$ in the
    range $10^{-5};10^{-3}$. The differential equation is valid only
    for times much larger than $t_c\approx8$, therefore we set
    $t_0=50$. As initial condition for the variable $y_0$ we used
    $y_0=y(t_0)=m_0t^2/2$, because the magnetization grows linearly
    during the initial regime. Red dashed lines represent $\pm
    m_c=\pm\lambda/(1-\lambda)$, also in this case disordered states
    are found in the region $[-m_c;+m_c]$.}
    \label{fig:ode_simulations}
\end{figure*}

The long time dynamics in the case $c \gg 1$ can be studied also in
more detail, by describing the evolution of the two peaks and their
mutual interaction. As shown in Appendix~\ref{other} we can write a
closed integro-differential equation for the magnetization $m(t)$. In
particular, defining
\[
y=\int_0^tm(t')dt'
\]
we obtain the following second order non linear ODE:
\begin{widetext}
\begin{equation}
  \frac{d^2y}{dt^2}=\sqrt{\frac{2}{\pi t (1-\lambda)^2}}
      \left\{
  \ton*{1-\frac{dy}{dt}}\frac{1}{\frac{1}{t}y-m_c}
  \exp\qua*{-\frac{t(1-\lambda)^2\ton*{\frac{1}{t}y-m_c}^2}{2}}-
  \ton*{1+\frac{dy}{dt}}\frac{1}{\frac{1}{t}y+m_c}
  \exp\qua*{-\frac{t(1-\lambda)^2\ton*{\frac{1}{t}y+m_c}^2}{2}}
  \right\}.
  \label{integrodiff}
\end{equation}
\end{widetext}
Note that the evolution of $m(t)$ is expressed in terms of
$y$, which is a variable containing the past history
of the system. This reflects the fact that personalized
information keeps memory of the preferences of the spin it is coupled to.
The fact that the evolution of $m$ is
governed by an integro-differential equation is then a very natural
consequence of the dynamics of the model. Solutions of this equation
are reported in Fig.~\ref{fig:ode_simulations}b, where it is possible to
see that states with $|m|<m_c$ remain disordered, while if $m<-m_c$ or
$m>m_c$ consensus is reached, in good agreement with numerical
simulations, shown in Fig.~\ref{fig:ode_simulations}a.

Disordered states for $c>1$ are completely different from the
disordered states for $c=1$. In the latter case the constant
magnetization is the effect of the external information
being a random uncorrelated variable equal for all nodes so that
all agents spend half of their time with $s_i=+1$  and half with
$s_i=-1$.
For $c>1$ instead, the system is divided in two polarized
clusters whose agents have a preferred spin value. A further
characterization in terms of self-overlap is presented in
Appendix~\ref{overlap}.

\section{Discussion and conclusions}

Let us summarize the results of our investigation.
The evolution of the system is described by the magnetization $m$
and the distribution of the local quantities $n_i$, describing the
personalized information for each agent.
Depending on whether $\delta$ ($c=1+\delta$) is much smaller or
much larger than $1$ the temporal evolution exhibits some variation.

In the first case there are three temporal regimes in the
evolution of the system.
For short times up to $2/\delta$ stochastic effects
dominate, the magnetization fluctuates around zero and
the distribution of the local $n_i$ remains centered around zero.
Later on the symmetry between positive and negative external information
breaks down because the 
mean value $n(t)$ of the single-peaked distribution $Q(n_i)$
starts drifting away from $n=0$ exponentially in time,
while also its width $\sigma(t)$ grows.
At the same time also $|m|$ grows exponentially.
This regime ends when linearization of the equations for
$m$ and $n$ is no more valid.
The nonlinear subsequent evolution varies depending on the
relative width of the peak. If the peak is narrow, all $n_i$ have
the same sign and consensus is quickly reached.
If the peak is broad then individuals with both $n_i>0$ and $n_i<0$ exist.
If the magnetization in this moment is small enough then
the system gets trapped in a disordered (polarized) state, where
disagreement persists and the magnetization keeps a constant value
$|m|<m_c = \lambda/(1-\lambda)$.

When $\delta \gg 1$ (i.e., $c \gg 1$) linearization
is never valid, the $Q(n_i)$ distribution always splits in two
components and this happens much earlier, over a time scale equal
to $t_c=1/\lambda^2$.
What happens next depends again on the value of the magnetization at $t_c$.
If $|m(t_c)|$ is sufficiently large, the drift of the two components has the
same sign. For example, if this sign is positive, it means that
individuals with negative $n_i$ have nevertheless an overall positive drift:
The negative component of the $Q(n_i)$ distribution gets rapidly
depleted and consensus is reached.
Otherwise the competition between the two opinions persists forever and
the system gets stuck in the polarized state with $m<m_c$.

\begin{figure}
  \includegraphics[width=0.49\textwidth]{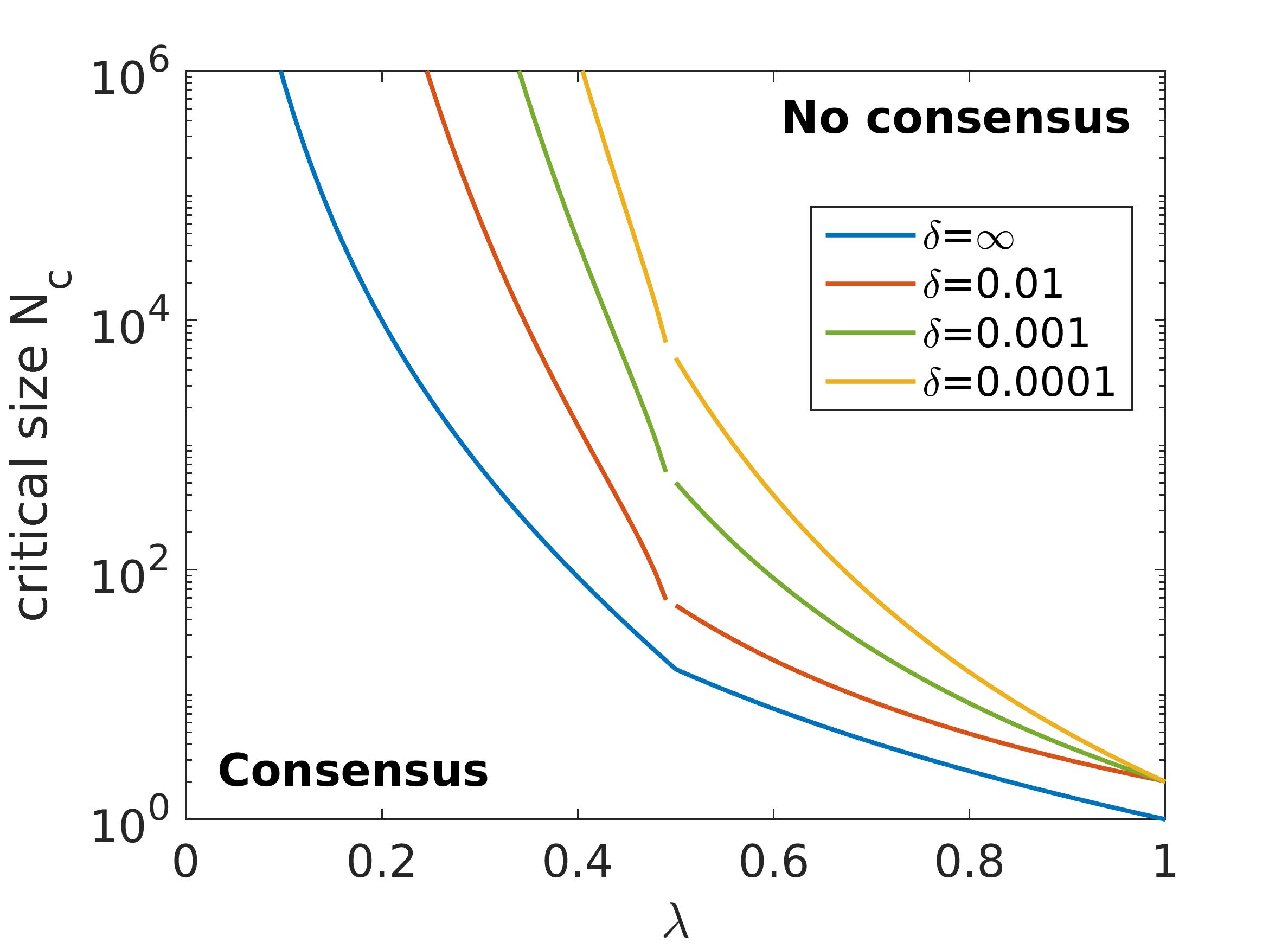}
  \caption{\textbf{Phase-diagram for all $c$.} The lines
  separate the region where consensus is reached
  from the region where it is not. Lines are obtained using
  Eq.~(\ref{N_csmallc}) for $\delta \ll 1$ and Eq.~(\ref{N_c2}) for
  $\delta=\infty$.}
  \label{pdallc}
\end{figure}

Although the detailed temporal evolution is rather different depending
on whether $c = 1+\delta$ is very close to 1 or much larger, the final overall
phenomenology is similar.  The parameter $c$ sets the temporal scales
of the dynamics and the details of the phase-diagram, but not the
qualitative features of the behavior: consensus for $N<N_c(\lambda)$,
polarization otherwise.
The dependence of $N_c$ on $\lambda$ is different depending on
whether $c$ is close to 1 [Eq.~(\ref{N_csmallc})] or large [Eq.~(\ref{N_c2})],
so the boundaries between the two regions depend on the value of $c$.
Fig.~\ref{pdallc} represents this phase-diagram for several of these
values, showing that, as expected, increasing the strength of personalized
information makes consensus more difficult.
However, we observe that increasing $c$ reduces the consensus region
but with a nontrivial limit: even in the $c \to \infty$ limit,
consensus is possible in sufficiently small systems.

This rather rich phenomenology has been obtained in the simplest
possible setting: an extremely simple binary opinion dynamics
always leading to consensus, in a mean-field framework,
augmented with an elementary form of personalized information.
Clearly our results open some interesting issues for future research:
What is the effect of a less trivial contact pattern among agents?
What changes when different opinion dynamics models are considered?
What happens if personalized information is parameterized differently, for
example considering different functional forms for the probability $P(n_i)$?
Finally, despite our oversimplified assumptions, some predictions
derived in this work may be testable in empirical systems. In particular,
the fact that disordered states cannot have magnetization larger than $m_c$
or the existence of a maximum size $N_c$ for the reaching of consensus
could be observable in systems where a community has to decide between
two alternative options.

%

\appendix

\section{Drift and diffusion coefficients}
\label{coefficients}
Let us define as $R_i$ the probability that $s_i$ switches
from $-1$ to $+1$ in a single update.
Eq.~\eqref{eq:evolution_s_i} implies
\[
R_i =  \frac{1}{N}\qua*{\frac{1-\lambda}{N}\ton*{\frac{1-s_i}{2}}\underbrace{\sum_j\ton*{\frac{1+s_j}{2}}}_{N_{\uparrow}}+\lambda\ton*{\frac{1-s_i}{2}} P(n_i)},
\]
where the prefactor stems from the fact that the $i$th spin is selected with probability $1/N$. Analogously $L_i$ is the probability of the opposite transition, namely from $s_i=+1$ to $s_i=-1$:
\[
L_i = \frac{1-\lambda}{N^2}\ton*{\frac{1+s_i}{2}}\underbrace{\sum_j\ton*{\frac{1-s_j}{2}}}_{N_{\downarrow}}+\frac{\lambda}{N} \ton*{\frac{1+s_i}{2}}\qua*{1-P(n_i)}.
\]
We can obtain the corresponding transition probabilities for the magnetization $m$ by summing over all spins 
\begin{eqnarray*}
  R^m 
  &=&(1-\lambda)\frac{1-m}{2}\cdot\frac{1+m}{2}+\frac{\lambda}{N}\sum_i \ton*{\frac{1-s_i}{2}}P(n_i),\\
  L^m 
  &=&(1-\lambda)\frac{1+m}{2}\cdot\frac{1-m}{2}+\frac{\lambda}{N}\sum_i \ton*{\frac{1+s_i}{2}}\qua*{1-P(n_i)}.
\end{eqnarray*}
In a single update, occurring in a time $\delta t=1/N$,
the variation of the magnetization is $\delta m = 2/N$, hence its drift is
\begin{eqnarray}
  v^m &=& \frac{\delta m}{\delta t} [R^m-L^m]\nonumber \\
  &=& 2 \frac{\lambda}{N}\sum_i \left\{ \left(\frac{1-s_i}{2} \right) P(n_i) - \left(\frac{1+s_i}{2} \right) [1-P(n_i)] \right\} \nonumber 
\end{eqnarray}			
Analogously the diffusion coefficient is
\begin{eqnarray}
  D^m &=& \frac{(\delta m)^2}{\delta t} [R^m+L^m]  \nonumber
  =  \frac{1-\lambda}{N}(1-m^2) \\
  &+&\frac{2\lambda}{N^2}
  \sum_i \left\{ \left(\frac{1-s_i}{2} \right) P(n_i) + \left(\frac{1+s_i}{2} \right) [1-P(n_i)] \right \}\nonumber
\end{eqnarray}

As expected, the evolution of the magnetization depends also on
the variables $n_i$, encoding the probability of the personalized
information. For this reason it is necessary to consider in detail
their evolution.
Their transition probabilities are
\begin{align}
  \label{eq:R^n_i}
  &R^{n_i} = \frac{1}{N}\qua*{(1-\lambda) \left(\frac{1+m}{2} \right) + \lambda P(n_i)},\\
  &L^{n_i} = \frac{1}{N}\qua*{(1-\lambda) \left(\frac{1-m}{2} \right) + \lambda [1-P(n_i)]}\nonumber.
\end{align}
The difference between these probabilities and those for the $s_i$
stems from the fact that $n_i$ is updated at each step, not only when
$s_i$ changes its state.
Using these expressions we can determine the drift for individual
variables $n_i$
\be
v_i = \frac{\delta n_i}{\delta t} (R^{n_i} - L^{n_i}) = (1-\lambda)m+ \lambda [2P(n_i)-1].
\ee

Summing the transition probabilities over all nodes we obtain the probabilities for $n$:
\begin{align*}
  &R^n =\sum_i R^{n_i} = (1-\lambda) \left(\frac{1+m}{2} \right)+ \frac{\lambda}{N} \sum_i P(n_i),\\
  &L^n = \sum_i L^{n_i} = (1-\lambda) \left(\frac{1-m}{2} \right)+ \frac{\lambda}{N} \sum_i [1-P(n_i)].
\end{align*}
In this case the variation in a single update is $\delta n=1/N$, as a
consequence the drift coefficient reads~
\be
v^n = \frac{\delta n}{\delta t} [R^n-L^n]=
(1-\lambda) m + \frac{\lambda}{N} \sum_i
\left[2 P(n_i)-1 \right].
\ee

The diffusion coefficient is instead
\begin{equation}
  D^n = \frac{(\delta n)^2}{\delta t} [R^n+L^n] =\frac{1}{N} \left[(1-\lambda) + \frac{\lambda}{N} \sum_i 1 \right] = \frac{1}{N}.
\end{equation}

\section{Variance of $Q(n_i)$ for $c \gtrsim 1$}
\label{Smallc}

The master equation governing the evolution of $Q(n_i)$ is
\begin{align*}
	Q(n_i, t+\delta t)=&R^{n_i-1} Q(n_i-1, t)+L^{n_i+1} Q(n_i+1, t)+\\
	&+\ton*{1-R^n-L^n}Q(n_i, t),
\end{align*}
  where $1-R^n-L^n=(1-1/N)$ is the probability that the spin $i$ is not selected
  and thus not updated. Expanding for small $\delta t$ we obtain
\begin{align}
  \frac{d}{dt} Q(n_i) = \frac{1}{\delta t}[&R^{n_i-1} Q(n_i-1, t) + L^{n_i+1} Q(n_i+1)\nonumber\\
  &-\ton*{R^n+L^n}Q(n_i)],
\end{align}
where $\delta t=1/N$ and the transition probabilities $R^{n_i}$ are given by
Eq.~\eqref{eq:R^n_i}. Noticing that $R^n+L^n=\frac{1}{N}$ we rewrite this equation as 
\begin{align}
  \frac{d}{dt} Q(n_i) =&NR^{n_i-1} Q(n_i-1, t) + \ton*{1-NR^{n_i+1}}Q(n_i+1)\nonumber\\
  &-Q(n_i),
  \label{eq:master_delta}
\end{align}
Using the condition $|n_i|\delta\ll 1$ we rewrite the transition probabilities as
\begin{equation}
  R^{n_i} 
   = \frac{1}{N}\qua*{(1-\lambda) \left(\frac{1+m}{2} \right) +\lambda \left(\frac{1}{2} + \frac{1}{4} \delta n_i \right)}.
\end{equation}
Substituting this expression into Eq.~\eqref{eq:master_delta},
multiplying by $n_i^2$ and summing over $n_i$ we obtain
\begin{eqnarray}
  \frac{d}{dt} \av{n_i^2}& = & \sum_{n_i} \left[ 4N R^{n_i} n_i Q(n_i) + (1-2n_i) Q(n_i)  \right]\nonumber \\
  & = & 2 m (1-\lambda) \av{n_i} + 1 + \lambda \delta \av{n_i^2},
  \label{eq:evolution_n^2_delta}
\end{eqnarray}
where $\av{n_i}$ is the quantity $n$ defined in Eq.~(\ref{n}).
The variance of $Q(n_i)$ is $\sigma^2 = \av{n_i^2} - \av{n_i}^2$, 
hence using Eqs.~\eqref{eq:system_m_n_delta} and \eqref{eq:evolution_n^2_delta} we get
\begin{align*}
  \frac{d}{dt} \sigma^2  = & \frac{d}{dt} \av{n_i^2} - 2 \av{n_i} \frac{d}{dt} \av{n_i} \\
  = & 2 m (1-\lambda) \av{n_i} + 1 + \lambda \delta \av{n_i^2}\\ 
  &- 2 \av{n_i}\left[(1-\lambda) m + \frac{\delta \lambda}{2} \av{n_i} \right] \\
  = & 1 + \lambda \delta \sigma^2,
\end{align*}
whose solution, with initial condition $\sigma^2(0)=0$, is
\begin{equation}
  \sigma^2(t) = \frac{1}{\lambda \delta} \left(\me^{\lambda \delta t} - 1\right).
\end{equation}

\section{A more refined lower bound for $N_c$}
\label{refined}
The drift of spin $i$ is given by Eq.~\eqref{eq:drift_v_i}
\[
v_i=m(1-\lambda)+\lambda\qua*{2P(n_i)-1}.
\]
For small $\delta=c-1$, the probability $P(n_i)$ can be expanded, giving  
\[
P(n_i)=\frac{1}{2}+\frac{n_i\delta}{4},
\]
therefore 
\[
v_i=m(1-\lambda)+\lambda\frac{n_i\delta}{2}.
\]	
As a consequence the spins with negative drift are those whose $n_i$
satisfies $n_i<\bar{n}$, with $\bar{n}$ defined by
\begin{equation}
  m(1-\lambda)+\lambda\frac{\bar{n}\delta}{2}=0\ \to\ \bar{n}=-\frac{2m}{\delta}\frac{1-\lambda}{\lambda}=-\frac{2}{\delta}\frac{m}{m_c}.
  \label{eq:bar_n}
\end{equation}
We assume that at time $T^*$, when linearization breaks down, all spins
with $n_i<0$, but positive drift $v_i(T^*)>0$,
will eventually end up with $n_i \to +\infty$,
while those with $v_i(T^*)<0$ will drift toward $-\infty$.
The number of spins with $v_i(T^*)<0$ is easily obtained by
integrating the distribution $Q_t(n_i)$ and using
Eq.~\eqref{eq:bar_n}.
More precisely, denoting this quantity by
$N^-_{\infty}$, it holds
\[
N^-_{\infty}=N\int_{-\infty}^{\bar{n}}Q(n_i)dn_i=N\int_{-\infty}^{-\frac{2}{\delta}\frac{m(T^*)}{m_c}}Q_{T^*}(n_i)dn_i.
\]
The distribution $Q_t(n_i)$ can be approximated by a Gaussian with mean value $n(t)$ and variance $\sigma^2(t)$. At time $T^*$ the variance satisfies by definition
\[
\sigma^2(T^*)=\frac{1}{\delta^2},
\]
while recalling that 
\[
n(t)=\sqrt{\frac{2}{\delta N}}\me^{t\delta/2}, \quad T^*=\frac{1}{\lambda\delta}\log\ton*{1+\frac{\lambda}{\delta}},
\]
we get 
\begin{equation}
  n(T^*)=\sqrt{\frac{2}{\delta N}}\ton*{1+\frac{\lambda}{\delta}}^{\frac{1}{2\lambda}}.
  \label{eq:n_T^*}
\end{equation}
Asymptotically, the number of negative spin can then be written as 
\begin{align*}
  &N^-_{\infty}=\frac{N}{\sqrt{2\pi\sigma^2(T^*)}}\int_{-\infty}^{-\frac{2}{\delta}\frac{m(T^*)}{m_c}}\exp\qua*{-\frac{\ton*{n(T^*)-n_i}^2}{2\sigma^2(T^*)}}dn_i=\\
  &=\frac{N\delta}{\sqrt{2\pi}}\int_{-\infty}^{-\frac{2}{\delta}\frac{m(T^*)}{m_c}}\exp\gra*{-\frac{\qua*{\sqrt{\frac{2}{\delta N}}\ton*{1+\frac{\lambda}{\delta}}^{\frac{1}{2\lambda}}-n_i}^2\delta^2}{2}}dn_i
\end{align*}
The asymptotic magnetization $m_{\infty}$ will consequently be 
\begin{align*}
  &m_{\infty}=\frac{N^+_{\infty}-N^-_{\infty}}{N}=\frac{N-2N^-_{\infty}}{N}=\\
  &=1-2\frac{\delta}{\sqrt{2\pi}}\int_{-\infty}^{-\frac{2}{\delta}\frac{m(T^*)}{m_c}}\exp\gra*{-\frac{\qua*{\sqrt{\frac{2}{\delta N}}\ton*{1+\frac{\lambda}{\delta}}^{\frac{1}{2\lambda}}-n_i}^2\delta^2}{2}}dn_i.
\end{align*}
The asymptotic state is disordered only if this magnetization is smaller
than the critical magnetization, otherwise the drift of the negative
peak is positive and therefore consensus is reached. The condition for
consensus consequently reads
\[
m_{\infty}>m_c=\frac{\lambda}{1-\lambda},
\]
that is
\begin{equation}
  1-2\frac{\delta}{\sqrt{2\pi}}\int_{-\infty}^{-\frac{2}{\delta}\frac{m(T^*)}{m_c}}\exp\gra*{-\frac{\qua*{\sqrt{\frac{2}{\delta N}}\ton*{1+\frac{\lambda}{\delta}}^{\frac{1}{2\lambda}}-n_i}^2\delta^2}{2}}dn_i>m_c.
  \label{eq:N_c_refined}
\end{equation}
Here, using Eq.~\eqref{eq:n_T^*} and recalling that $m=\frac{\delta}{2}n$, $m(T^*)$ satisfies
\[
m(T^*)=\sqrt{\frac{\delta}{2N}}\ton*{1+\frac{\lambda}{\delta}}^{\frac{1}{2\lambda}}.
\]
By solving numerically Eq.~\eqref{eq:N_c_refined} for the value of $N$,
we derive a numerical lower bound of the critical size $N_c$ of systems
reaching consensus.

\section{Variance of $Q(n_i)$ for $c \gg 1$}
\label{Largec}

We can study the evolution of the component $Q^+(n_i)$ using the master
equation defined in Eq.~\eqref{eq:master_delta}
\begin{align}
  \frac{d}{dt} Q^+(n_i) = &NR^{n_i-1} Q^+(n_i-1) +\nonumber\\
  &+\ton*{1-N R^{n_i+1}} Q^+(n_i+1) - Q^+(n_i),
  \label{eq:master_equation}
\end{align}
where the transition probabilities $R^{n_i}$ satisfy
\[
R^{n_i}=\frac{1}{N}\qua*{(1-\lambda)\frac{1+m}{2}+\lambda P(n_i)}.
\]
The first moment $n^+$ varies as 
\begin{align*}
  v^+=&\frac{dn^+}{dt}=\sum_{n_i}\frac{d}{dt} Q^+(n_i)n_i=\\
  =&\sum_{n_i}Q^+(n_i)\qua*{2NR^{n_i}-1}.
\end{align*}
Assuming $Q^+(n_i\leq0) \approx 0$,
we can rewrite the transition probabilities as 
\begin{equation}
  N\cdot R^{n_i}=(1-\lambda)\frac{1+m}{2}+\lambda\ton*{1-c^{-n_i}},
  \label{eq:transition_positive}
\end{equation}
and then obtain for the drift $v^+$
\begin{align}
  v^+=&\sum_{n_i}Q^+(n_i)\qua*{2\lambda+(1-\lambda)(1+m)-2\lambda c^{-n_i}-1}=\nonumber\\
  =&m(1-\lambda)+\lambda-2\lambda c^{-n^+}\approx m(1-\lambda)+\lambda,
  \label{eq:nu_plus}
\end{align}
where we have made the approximation $\mean*{f(n)}=f\ton*{\mean{n}}$.
Similarly 
\begin{equation}
  v^-=m(1-\lambda)-\lambda+2\lambda c^{n^-}\approx m(1-\lambda)-\lambda.
  \label{eq:nu_minus}
\end{equation}

Let us now turn to the evolution of the variance.
As shown in the derivation
of Eq.~\eqref{eq:evolution_n^2_delta} it holds
\[
\frac{d\mean*{n^2}_+}{dt}=\sum_{n_i}\qua*{4NR^{n_i}Q^+(n_i)n_i+(1-2n_i)Q^+(n_i)}.
\]
Again, making the approximation $Q^+(n_i\leq 0)\approx 0$
we can use Eq.~\eqref{eq:transition_positive}
\begin{align*}
  \frac{d\mean*{n^2}_+}{dt}=\sum_{n_i}&\left\{4\qua*{(1-\lambda)\frac{1+m}{2}+\lambda}Q^+(n_i)n_i\right.\\
  &\hspace{1em}+(1-2n_i)Q^+(n_i)\bigg\},
\end{align*}
hence
\[
\frac{d\mean*{n^2}_+}{dt}=2\qua*{\lambda+m(1-\lambda)}n^++1.
\]
This implies that the variance $\sigma^2_+(t)$ of $Q^+(n_i)$ evolves as 
\begin{align*}
  \frac{d\sigma^2_+}{dt}&=\frac{d\mean*{n^2}_+}{dt}-2n^+\frac{d}{dt}n^+=\frac{d\mean*{n^2}_+}{dt}-2n^+v^+\\
  &=2\qua*{\lambda+m(1-\lambda)}n^++1\\
  &\hspace{1em}-2n^+\qua*{m(1-\lambda)+\lambda}=1.
\end{align*}
Assuming $\sigma^2_+(0)=0$ we obtain
\begin{equation}
  \sigma^2_+(t)=t.
  \label{eq:sigma+}
\end{equation}
Similarly $\sigma^2_-(t)=t$.
We then conclude that the two peaks widen
as it would happen for an unbiased random walk.

We can now study the dynamics of the variance $\sigma^2(t)$ of
$Q(n_i)$. For a bimodal distribution it holds
\[
\sigma^2=p\qua*{\sigma_1^2+(\mu_1-\mu)^2}+(1-p)\qua*{\sigma_2^2+(\mu_2-\mu)^2},
\]
that in our case becomes 
\begin{align*}
  \sigma^2&=\frac{N^+}{N}\qua*{\sigma_+^2+(n_+-n)^2}+\frac{N^-}{N}\qua*{\sigma_-^2+(n_--n)^2}\\
  &=\frac{N^+}{N}\gra*{\sigma_+^2+\qua*{\frac{N^-}{N}(n^+-n^-)}^2}+\\
  &\hspace{1em}+\frac{N^-}{N}\gra*{\sigma_-^2+\qua*{\frac{N^+}{N}(n^+-n^-)}^2}.
\end{align*}
The quantities $N^+$ and $N^-$ vary slowly, we can then approximate them as constants and derive over time, obtaining
\begin{align*}
  \frac{d\sigma^2}{dt}=&\frac{N^+}{N}\gra*{\frac{d\sigma_+^2}{dt}+\ton*{\frac{N^-}{N}}^22(n^+-n^-)\frac{d(n^+-n^-)}{dt}}\\
  &+\frac{N^-}{N}\gra*{\frac{d\sigma_-^2}{dt}+\ton*{\frac{N^+}{N}}^22(n^+-n^-)\frac{d(n^+-n^-)}{dt}}=\\
  =&\frac{N^+}{N}\qua*{1+\ton*{\frac{N^-}{N}}^22(v^+-v^-)^2t}+\\
  &+\frac{N^-}{N}\qua*{1+\ton*{\frac{N^+}{N}}^22(v^+-v^-)^2t},
\end{align*}
where we used Eq.~\eqref{eq:sigma+}. Noting that
Eqs.~\eqref{eq:nu_plus} and \eqref{eq:nu_minus} imply
$v^+-v^-\approx2\lambda$ and recalling that
$c\gg1$ we obtain
\[
\frac{d\sigma^2}{dt}\approx1+8\frac{N^+N^-}{N^2}\lambda^2t
\]
It is easy to show that $N^+$ and $N^-$ can be approximated by
$N_{\uparrow}$ and $N_{\downarrow}$. Indeed using
Eq.~\eqref{eq:drift_m_general} we can write the drift for $m$ as
\[
v^m=\frac{\lambda}{N} \sum_i \left(\frac{c^{n_i}-1}{c^{n_i}+1}\right)-\lambda m \approx \lambda\qua*{\frac{N^+-N^-}{N}-m},
\]
which implies that, up to diffusive contributions,
$m\approx \frac{N^+-N^-}{N}$.
We can then set $N^+\approx N(1+m)/2$ arriving at the following expression
\begin{equation}
  \frac{d\sigma^2}{dt}=1+2(1-m^2)\lambda^2t.
  \label{eq:evolution_sigma}
\end{equation}
For small $m$ we can neglect the term $m^2$ obtaining 
\[
\frac{d\sigma^2}{dt}\approx1+2\lambda^2t.
\]
which, setting $\sigma^2(0)=0$, yields
\begin{equation}
  \sigma^2(t)=t+\lambda^2t^2.
\end{equation}

\section{Integro-differential equation for $m$}
\label{other}
Up to $t=t_c$ the distribution $Q(n_i)$ is unimodal and can be
described in terms of $n$ and $\sigma^2$; to understand the subsequent
evolution, we approximate the distribution $Q(n_i)$ as the sum of two
Gaussian distributions $Q^+(n_i)$ and $Q^-(n_i)$ with variance
$\sigma^2_\pm=t$ [See Eq.~\eqref{eq:sigma+}] and mean values $n^+$ and
$n^-$. Let us focus on the $Q^+(n_i)$ component, this peak moves with
drift $v^+$ and widens linearly in time. This implies that,
in the reference frame moving with the peak, each spin performs
and unbiased random walk. As a consequence, neglecting temporal
correlations, the probability that one of these spins during the random
walk reaches negative values of $n_i$ can be obtained by integrating
$Q^+(n_i)$ for all $n_i<0$. The probability $P_{R\to L}$ that an
individual performs a transition from $n_i>0$ to $n_i<0$ is then:
\begin{align} \nonumber
  P_{R\to L}&=\int_{-\infty}^0Q^+(n)dn\approx\frac{1}{\sqrt{2\pi\sigma_+^2}}\int_{-\infty}^0\exp\qua*{-\frac{(n-n^+)^2}{2\sigma_+^2}}=\\ 
  &=\frac{1}{2}\qua*{1-\erf\ton*{\frac{n^+}{\sqrt{2\sigma_+^2}}}}
  \label{erf}
\end{align}
Focusing on $t\gg t_c$ we can assume that the two peaks are well separated
and therefore we can expand the error function for large argument
\[
\erf(x)\approx1-\frac{1}{\sqrt{\pi}x}\exp\ton*{-x^2}
\]
obtaining
\begin{equation}
  P_{R\to L}(t)\approx\sqrt{\frac{2}{\pi}}\frac{\sigma_+(t)}{n^+(t)}\exp\qua*{-\frac{\ton*{n^+(t)}^2}{2\sigma_+^2(t)}}.
  \label{eq:P_RL_initial}
\end{equation}
We can express the mean value $n^+(t)$ in terms of the drift $v^+(t)$ as 
\[
n^+(t)=\int_0^tv^+(t')dt'=\lambda t+(1-\lambda)\int_0^tm(t')dt',
\]
where we used Eq.~\eqref{eq:nu_plus}. Substituting this expression
into Eq.~\eqref{eq:P_RL_initial} we get
\begin{align*}
  P_{R\to L}(t)\approx&\sqrt{\frac{2}{\pi}}\frac{\sigma_+(t)}{\lambda t+(1-\lambda)\int_0^tm(t')dt'}\times\\
  &\times\exp\gra*{-\frac{\qua*{\lambda t+(1-\lambda)\int_0^tm(t')dt'}^2}{2\sigma_+^2(t)}}.
\end{align*}

Recalling that $\sigma_+(t)=t$, defining
$m_c=\frac{\lambda}{1-\lambda}$ and introducing the temporal average
of the magnetization
\be
\bar{m}(t)=\frac{1}{t}\int_0^tm(t')dt'
\ee
we can rewrite this probability as
\begin{equation}
  P_{R\to L}(t)\approx\sqrt{\frac{\tau^+(t)}{\pi t}}\exp\ton*{-\frac{t}{\tau^+(t)}},
  \label{eq:transition_prob_positive}
\end{equation}
where we introduced the characteristic temporal scale
$\tau^+(t)$ which satisfies 
\begin{equation}
  \tau^+(t)=\frac{2}{(1-\lambda)^2}\frac{1}{\qua*{\bar{m}(t)+m_c}^2}.
  \label{eq:tau_positive}
\end{equation}

Of course perfectly analogous formulas hold for the negative peak:
\begin{equation}
  P_{L\to R}(t)\approx\sqrt{\frac{\tau^-(t)}{\pi t}}\exp\ton*{-\frac{t}{\tau^-(t)}},
  \label{eq:transition_prob_negative}
\end{equation}	
where the characteristic time $\tau^-(t)$ satisfies 
\begin{equation}
  \tau^-(t)=\frac{2}{(1-\lambda)^2}\frac{1}{\qua*{\bar{m}(t)-m_c}^2}.
  \label{eq:tau_negative}
\end{equation}

Eq.~\eqref{eq:transition_prob_positive} implies that
the transition probability is exponentially small for $t>\tau^+(t)$,
while if $t<\tau^+(t)$ the spins can make a
transition from $n_i>0$ to $n_i<0$ with non vanishing probability.
If $\tau^+\to\infty$ the transition probability in
Eq.~\eqref{eq:transition_prob_positive} becomes essentially constant.
This is a manifestation of the fact that the expansion of
the error function in Eq.~\eqref{erf} cannot be performed as
the positive peak is not narrow. Physically, this
implies that the positive peak gets rapidly absorbed by the
negative one leading to consensus.
In order to determine if consensus is reached
we then have to study this characteristic temporal scale.

We first observe that if, starting from a given time,
the magnetization is smaller than $-m_c$ then
$\bar{m}(t)$, which starts from $\bar{m}(t)=0$,
gradually decreases, necessarily reaching at some point the value $-m_c$.
Hence $\tau^+$ diverges and consensus is rapidly reached.
If instead $m$ does not become smaller than $-m_c$ then $\bar{m}(t)$
always remains larger than $-m_c$, $\tau^+$ remains finite and
consensus is never reached.
Hence we can predict that asymptotically disordered configurations
are possible only for magnetization in the interval $-m_c < m < m_c$,
while if $m$ exceeds these bounds consensus is reached.
Fig.~\ref{fig:ode_simulations} confirms this expectation.

We can use the transition probabilities for writing a closed-form
equation for $m(t)$.
The derivative of $N^+$ satisfies
\[
\frac{dN^+}{dt}=N^-(t)P_{L\to R}(t)-N^+(t)P_{R\to L}(t).
\]
Expressing $N^+$ and $N^-$ in terms of $m$ we get
\be
\frac{dm}{dt}=\qua*{1-m(t)}P_{L\to R}(t)-\qua*{1+m(t)}P_{R\to L}(t).
\label{dmdt}
\ee
Inserting
Eqs.~(\ref{eq:transition_prob_positive}-\ref{eq:tau_negative})
into Eq.~\eqref{dmdt} we finally
obtain an integro-differential equation for the magnetization $m(t)$,
\begin{align*}
  	\frac{dm}{dt}=&\qua*{1-m(t)}\sqrt{\frac{2}{\pi t\qua*{\frac{1}{t}\int_0^tm(t')dt'-m_c}^2(1-\lambda)^2}}\times\\
	&\times\exp\gra*{-\frac{t(1-\lambda)^2\qua*{\frac{1}{t}\int_0^tm(t')dt'-m_c}^2}{2}}-\\
	&-\qua*{1+m(t)}\sqrt{\frac{2}{\pi t\qua*{\frac{1}{t}\int_0^tm(t')dt'+m_c}^2(1-\lambda)^2}}\times\\
	&\times\exp\gra*{-\frac{t(1-\lambda)^2\qua*{\frac{1}{t}\int_0^tm(t')dt'+m_c}^2}{2}}.
\end{align*}

\section{Detecting filter-bubbles: self overlap}
\label{overlap}
In order to properly characterize disordered configurations,
characterized by $|m|<1$, we consider the local magnetization
$m_i=\mean{s_i}$, defined as
\[
m_i(t)=P_+(t)\cdot 1+P_-(t)\cdot (-1),
\]
where 
\begin{align*}
  P_+=&(1-\lambda)\frac{N_{\uparrow}(t)}{N}+\lambda P(n_i(t)),\\
  P_-=&(1-\lambda)\frac{N_{\downarrow}(t)}{N}+\lambda [1-P(n_i(t))].
\end{align*}
Noting that the mean magnetization $m(t)=1/N\sum m_i(t)$ satisfies
\[
m(t)=\frac{N_{\uparrow}(t)-N_{\downarrow}(t)}{N} \ \to \ N_{\uparrow}(t)=\frac{N\qua*{m(t)+1}}{2},
\]
we obtain
\begin{align*}
  m_i(t)=&(1-\lambda)\frac{1+m(t)}{2}+\lambda P(n_i(t))-\\
  &-(1-\lambda)\frac{1-m(t)}{2}-\lambda [1-P(n_i(t))].
\end{align*}
Therefore the magnetization of site $i$ is 
\begin{equation}
  m_i(t)=m(t)(1-\lambda)+\lambda\qua*{2P(n_i(t))-1}.
  \label{eq:magnetization_site}
\end{equation}
We can now introduce the self overlap $q(t)$ which allows to detect
the presence of filter bubbles
\begin{equation}
  q(t)=\frac{1}{N}\sum_i^N m_i(t)^2.
  \label{eq:overlap}
\end{equation}
Indeed if the network is split in two bubbles this quantity is
expected to be different from zero even if the overall magnetization
is null, because each node tends to be aligned to one of the two
opinions. It is easy to show that the disordered state found for $c=1$
does not contain separate filter bubbles. Indeed using
Eq.~\eqref{eq:magnetization_site} and recalling that $P(n_i(t))=1/2$
we obtain
\[
q_{c=1}(t)=\frac{1}{N}\sum_i m(t)^2(1-\lambda)^2=m(t)^2(1-\lambda)^2\sim\frac{1}{\sqrt{N}},
\]
where we used the fact that the diffusion coefficient, defined by
Eq.~\eqref{eq:nu_D_c=1}, scales as $1/N$. We thus see that in the
limit of large systems the self overlap is null, meaning that each
node randomly flips between $+1$ and $-1$: no polarization is
present. This also implies that the variables $n_i$ perform an
unbiased random walk and therefore the width of the probability distribution
of the $n_i$, $Q(n_i)$, is expected to broaden as $\sqrt{t}$.
			
Let us turn now to the case $c>1$. As we have shown there are disordered
states in which the external information, after an initial transient,
completely polarizes. This implies that $N^+$ spins receive positive
information, while negative information acts on the remaining $N^-$ spins.
Let us
consider a positively polarized spin $i$, for which it holds
$P(n_i)=1$:  using expression \eqref{eq:magnetization_site} we obtain
for its magnetization
\[
m_i^+(t)=m(t)(1-\lambda)+\lambda,
\]
similarly for a negatively polarized spin it holds
\[
m_i^-(t)=m(t)(1-\lambda)-\lambda.
\]
Splitting the sum of relation \eqref{eq:overlap} we can write the overlap as
\begin{align*}
  q_{c>1}&=\frac{1}{N}\qua*{\sum_{i|n_i>0}\qua*{m(t)(1-\lambda)+\lambda}^2+\sum_{i|n_i<0}\qua*{m(t)(1-\lambda)-\lambda}^2}\\
  &=m(t)^2(1-\lambda)^2+\lambda^2+2m(t)^2\lambda(1-\lambda)
\end{align*}
In this case the overlap is non null also if $m=0$, meaning that the
variables are partially frozen due to the personalized
information. The disordered state is then profoundly different from
the one characteristic of $c=1$. Indeed a null overlap implies that
spins randomly flip, spending half of their time with a positive
orientation and half with a negative one. Conversely an overlap
different from zero indicates the presence of two polarized clusters,
whose spins tend to remain fixed.

\end{document}